\documentclass{eptcs}

\newif\ifignore 
\ignorefalse
\newcommand{\auxproof}[1]{
\ifignore\mbox{}\newline
\textbf{PROOF:} \dotfill\newline
{\it #1}\mbox{}\newline
\textbf{ENDPROOF}\dotfill
\fi}

\usepackage{amsmath}
\usepackage{amssymb}
\usepackage{amstext}
\usepackage{amsthm}
\usepackage{mathtools}
\usepackage{mathrsfs}
\usepackage{stmaryrd}
\usepackage{xspace}
\usepackage{enumerate}
\usepackage{url}
\usepackage{nicefrac}
\usepackage{fancybox}
\usepackage{hyperref}

\usepackage{xypic}
\usepackage[all,cmtip]{xy}
\xyoption{2cell}
\UseTwocells
\xyoption{tips}

\usepackage{tikz}
\usetikzlibrary{circuits.ee.IEC}

\newlength\stateheight
\newlength\minimumstatewidth
\tikzset{width/.initial=\minimummorphismwidth}
\tikzset{colour/.initial=white}

\newif\ifblack\pgfkeys{/tikz/black/.is if=black}
\newif\ifwedge\pgfkeys{/tikz/wedge/.is if=wedge}
\newif\ifvflip\pgfkeys{/tikz/vflip/.is if=vflip}
\newif\ifhflip\pgfkeys{/tikz/hflip/.is if=hflip}
\newif\ifhvflip\pgfkeys{/tikz/hvflip/.is if=hvflip}

\pgfdeclarelayer{foreground}
\pgfdeclarelayer{background}
\pgfsetlayers{background,main,foreground}

\def\thickness{0.4pt}

\makeatletter
\pgfkeys{%
  /tikz/on layer/.code={
    \pgfonlayer{#1}\begingroup
    \aftergroup\endpgfonlayer
    \aftergroup\endgroup
  },
  /tikz/node on layer/.code={
    \gdef\node@@on@layer{%
      \setbox\tikz@tempbox=\hbox\bgroup\pgfonlayer{#1}\unhbox\tikz@tempbox\endpgfonlayer\pgfsetlinewidth{\thickness}\egroup}
    \aftergroup\node@on@layer
  },
  /tikz/end node on layer/.code={
    \endpgfonlayer\endgroup\endgroup
  }
}
\def\node@on@layer{\aftergroup\node@@on@layer}
\makeatother

\makeatletter
\pgfdeclareshape{state}
{
    \savedanchor\centerpoint
    {
        \pgf@x=0pt
        \pgf@y=0pt
    }
    \anchor{center}{\centerpoint}
    \anchorborder{\centerpoint}
    \saveddimen\overallwidth
    {
        \pgf@x=3\wd\pgfnodeparttextbox
        \ifdim\pgf@x<\minimumstatewidth
            \pgf@x=\minimumstatewidth
        \fi
    }
    \savedanchor{\upperrightcorner}
    {
        \pgf@x=.5\wd\pgfnodeparttextbox
        \pgf@y=.5\ht\pgfnodeparttextbox
        \advance\pgf@y by -.5\dp\pgfnodeparttextbox
    }
    \anchor{A}
    {
        \pgf@x=-\overallwidth
        \divide\pgf@x by 4
        \pgf@y=0pt
    }
    \anchor{B}
    {
        \pgf@x=\overallwidth
        \divide\pgf@x by 4
        \pgf@y=0pt
    }
    \anchor{text}
    {
        \upperrightcorner
        \pgf@x=-\pgf@x
        \ifhflip
            \pgf@y=-\pgf@y
            \advance\pgf@y by 0.4\stateheight
        \else
            \pgf@y=-\pgf@y
            \advance\pgf@y by -0.4\stateheight
        \fi
    }
    \backgroundpath
    {
       \begin{pgfonlayer}{foreground}
        \pgfsetstrokecolor{black}
        \pgfsetlinewidth{\thickness}
        \pgfpathmoveto{\pgfpoint{-0.5*\overallwidth}{0}}
        \pgfpathlineto{\pgfpoint{0.5*\overallwidth}{0}}
        \ifhflip
            \pgfpathlineto{\pgfpoint{0}{\stateheight}}
        \else
            \pgfpathlineto{\pgfpoint{0}{-\stateheight}}
        \fi
        \pgfpathclose
        \ifblack
            \pgfsetfillcolor{black!50}
            \pgfusepath{fill,stroke}
        \else
            \pgfusepath{stroke}
        \fi
       \end{pgfonlayer}
    }
}

\pgfdeclareshape{my ground}
{
  \inheritsavedanchors[from=rectangle ee]
  \inheritanchor[from=rectangle ee]{center}
  \inheritanchor[from=rectangle ee]{north}
  \inheritanchor[from=rectangle ee]{south}
  \inheritanchor[from=rectangle ee]{east}
  \inheritanchor[from=rectangle ee]{west}
  \inheritanchor[from=rectangle ee]{north east}
  \inheritanchor[from=rectangle ee]{north west}
  \inheritanchor[from=rectangle ee]{south east}
  \inheritanchor[from=rectangle ee]{south west}
  \inheritanchor[from=rectangle ee]{input}
  \inheritanchor[from=rectangle ee]{output}
  \anchorborder{\csname pgf@anchor@my ground@input\endcsname}

  \backgroundpath{
    \pgf@process{\pgfpointadd{\southwest}{\pgfpoint{\pgfkeysvalueof{/pgf/outer xsep}}{\pgfkeysvalueof{/pgf/outer ysep}}}}
    \pgf@xa=\pgf@x \pgf@ya=\pgf@y
    \pgf@process{\pgfpointadd{\northeast}{\pgfpointscale{-1}{\pgfpoint{\pgfkeysvalueof{/pgf/outer xsep}}{\pgfkeysvalueof{/pgf/outer ysep}}}}}
    \pgf@xb=\pgf@x \pgf@yb=\pgf@y
    \pgfpathmoveto{\pgfqpoint{\pgf@xa}{\pgf@ya}}
    \pgfpathlineto{\pgfqpoint{\pgf@xa}{\pgf@yb}}
    \pgfmathsetlength\pgf@xa{.5\pgf@xa+.5\pgf@xb}
    \pgfmathsetlength\pgf@yc{.16666\pgf@yb-.16666\pgf@ya}
    \advance\pgf@ya by\pgf@yc
    \advance\pgf@yb by-\pgf@yc
    \pgfpathmoveto{\pgfqpoint{\pgf@xa}{\pgf@ya}}
    \pgfpathlineto{\pgfqpoint{\pgf@xa}{\pgf@yb}}
    \advance\pgf@ya by\pgf@yc
    \advance\pgf@yb by-\pgf@yc
    \pgfpathmoveto{\pgfqpoint{\pgf@xb}{\pgf@ya}}
    \pgfpathlineto{\pgfqpoint{\pgf@xb}{\pgf@yb}}
  }
}
\makeatother

\tikzset{inline text/.style =
  {text height=1.2ex, text depth=0.25ex,yshift=0.5mm}}
\tikzset{arrow box/.style =
  {rectangle,inline text,fill=white,draw,
    minimum height=5mm,yshift=-0.5mm,minimum width=5mm}}
\tikzset{dot/.style =
  {inner sep=0mm,minimum width=1mm,minimum height=1mm,
    draw,shape=circle}}
\tikzset{white dot/.style = {dot,fill=white,text depth=-0.2mm}}
\tikzset{scalar/.style = {diamond,draw,inner sep=1pt}}

\tikzset{copier/.style = {dot,fill,text depth=-0.2mm}}
\tikzset{discarder/.style = {my ground,draw,inner sep=0pt,
    minimum width=4.2pt,minimum height=11.2pt,anchor=input,rotate=90}}

\tikzset{uniform/.style = {my ground,draw,inner sep=0pt,
    minimum width=4.2pt,minimum height=11.2pt,anchor=input,rotate=270}}

\tikzset{xshiftu/.style = {shift = {(#1, 0)}}}
\tikzset{yshiftu/.style = {shift = {(0, #1)}}}

\newenvironment{myproof}{\begin{trivlist} \item[\hskip \labelsep%
{\bf Proof.}]}{\end{trivlist}}

\newcommand{\QEDbox}{\square}
\newcommand{\QED}{\hspace*{\fill}$\QEDbox$}

\DeclareSymbolFont{T1op}{T1}{cmr}{m}{n}
\SetSymbolFont{T1op}{bold}{T1}{cmr}{bx}{n}
\DeclareMathSymbol{\mathguilsinglleft}{\mathopen}{T1op}{'016}
\DeclareMathSymbol{\mathguilsinglright}{\mathclose}{T1op}{'017}

\newcommand{\idmap}[1][]{\ensuremath{\mathrm{id}_{#1}}}
\newcommand{\after}{\mathrel{\circ}}
\newcommand{\acc}{\ensuremath{\mathsl{acc}}}
\newcommand{\accs}{\ensuremath{\mathsl{accs}}}
\newcommand{\arr}{\ensuremath{\mathsl{arr}}}
\newcommand{\perm}{\ensuremath{\mathsl{perm}}}
\newcommand{\flrn}{\ensuremath{\mathsl{Flrn}}}
\newcommand{\zip}{\ensuremath{\mathsl{zip}}}
\newcommand{\mzip}{\ensuremath{\mathsl{mzip}}}
\newcommand{\lsplit}{\ensuremath{\mathsl{lsplit}}}
\newcommand{\msplit}{\ensuremath{\mathsl{msplit}}}
\newcommand{\concat}{\ensuremath{\mathbin{+{\kern-.5ex}+}}}
\newcommand{\del}{\ensuremath{\mathsl{del}}}

\newcommand{\drawdelete}{\ensuremath{\mathsl{DD}}}

\newcommand{\mulnom}{\ensuremath{\mathsl{mn}}}
\newcommand{\multinomial}[1][]{\ensuremath{\mulnom[#1]}}
\newcommand{\hypgeom}{\ensuremath{\mathsl{hg}}}
\newcommand{\hypergeometric}[1][]{\ensuremath{\hypgeom[#1]}}
\newcommand{\pml}{\ensuremath{\mathsl{pml}}}

\newcommand{\st}{\ensuremath{\mathsl{st}}}
\newcommand{\mst}{\ensuremath{\mathsl{mst}}}

\newcommand{\ket}[1]{\ensuremath{|{\kern.1em}#1{\kern.1em}\rangle}}
\newcommand{\bigket}[1]{\ensuremath{\big|{\kern.1em}#1{\kern.1em}\big\rangle}}

\newcommand{\distributionsymbol}{\mathcal{D}}

\newcommand{\multisetsymbol}{\mathcal{M}}

\newcommand{\Dst}{\distributionsymbol}

\newcommand{\Mlt}{\multisetsymbol}

\newcommand{\Giry}{\mathcal{G}}

\newcommand{\UF}{\ensuremath{\mathcal{U}{\kern-.75ex}\mathcal{F}}}

\newcommand{\Cat}[1]{\ensuremath{\mathbf{#1}}\xspace}
\newcommand{\cat}[1]{\Cat{#1}}
\newcommand{\Kl}{\mathcal{K}{\kern-.4ex}\ell}
\newcommand{\EM}{\mathcal{E}{\kern-.4ex}\mathcal{M}}

\newcommand{\NNO}{\mathbb{N}}

\newcommand{\Ef}{\ensuremath{\mathcal{E}{\kern-.5ex}f}}
\newcommand{\intd}{{\kern.2em}\mathrm{d}{\kern.03em}}

\newcommand{\OF}{\ensuremath{\mathcal{O}{\kern-.1em}\mathcal{F}}}
\newcommand{\Closed}{\ensuremath{\mathcal{C}{\kern-.05em}\ell}}

\newcommand{\congrightarrow}{\mathrel{\smash{\stackrel{
           \raisebox{.5ex}{$\scriptstyle\cong$}}{
           \raisebox{0ex}[0ex][0ex]{$\rightarrow$}}}}}

\newsavebox\sbpto
\savebox\sbpto{\begin{tikzpicture}[baseline=-2.5pt]
            \filldraw[fill=white,draw=white] circle (1.4pt);
            \filldraw[fill=white,draw=black,line width=0.2pt] circle
(1.2pt);
                \end{tikzpicture}}

\makeatother
 \newdir{ >}{{}*!/-7.5pt/@{>}}
 \newdir{|>}{!/4.5pt/@{|}*:(1,-.2)@^{>}*:(1,+.2)@_{>}}
 \newdir{ |>}{{}*!/-3pt/@{|}*!/-7.5pt/:(1,-.2)@^{>}*!/-7.5pt/:(1,+.2)@_{>}}

\newsavebox\sbground
\savebox\sbground{\begin{tikzpicture}[circuit ee IEC,yscale=0.5,xscale=0.4]
                \draw (0,-2ex) to (0,0) node[ground,rotate=90,xshift=.65ex] {};
                \end{tikzpicture}}

\newsavebox\sbunif
\savebox\sbunif{\begin{tikzpicture}[circuit ee IEC,yscale=0.6,xscale=0.5]
                \draw (0,0) to (0,2ex) node[ground,rotate=270,xshift=2.5ex] {};
                \end{tikzpicture}}
\newcommand\unif{\mathbin{\text{\raisebox{-0.3ex}{\usebox\sbunif}}}}

\ifdefined\mathsl\else
  \DeclareMathAlphabet{\mathsl}{\encodingdefault}{\rmdefault}{\mddefault}{\sldefault}
  \SetMathAlphabet{\mathsl}{bold}{\encodingdefault}{\rmdefault}{\bfdefault}{\sldefault}
\fi

\makeatletter
\newcommand{\mathoverlap}[2]{\mathpalette\mathoverlap@{{#1}{#2}}}
\newcommand{\mathoverlap@}[2]{\mathoverlap@@{#1}#2}
\newcommand{\mathoverlap@@}[3]{\ooalign{$\m@th#1#2$\crcr\hidewidth$\m@th#1#3$\hidewidth}}
\makeatother

\theoremstyle{plain}
\newtheorem{theorem}{Theorem}
\newtheorem{lemma}{Lemma}
\newtheorem{proposition}[lemma]{Proposition}

\theoremstyle{definition}
\newtheorem{definition}{Definition}

\providecommand{\event}{MFPS 2021}
\title{Multinomial and Hypergeometric Distributions \\ in Markov Categories}

\author{Bart Jacobs\institute{Institute
    for Computing and Information Sciences (iCIS), 
\\ Radboud  University Nijmegen, The Netherlands.
}  
\email{bart@cs.ru.nl} 
}

\def\titlerunning{Distributions in Markov Categories}
\def\authorrunning{B. Jacobs}

\date{\small \today}

\begin{document}
\maketitle

\begin{abstract} 
Markov categories, having tensors with copying and discarding, provide
a setting for categorical probability. This paper uses finite colimits
and what we call uniform states in such Markov categories to define a
(fixed size) multiset functor, with basic operations for sums and zips
of multisets, and a graded monad structure. Multisets can be used to
represent both urns filled with coloured balls and also draws of
multiple balls from such urns.  The main contribution of this paper is
the abstract definition of multinomial and hypergeometric
distributions on multisets, as draws.  It is shown that these
operations interact appropriately with various operations on
multisets.
\end{abstract}




\section{Introduction}\label{IntroSec}

Given the current reliance on the probabilistic analysis of huge
datasets, it is important to have a good formal understanding of what
may be called the logic of probability. In this line of work there is
growing interest in the axiomatisation of probability theory, using
\textit{e.g.}~category theory --- also called `synthetic' probability
theory. Several efforts and approaches can be distinguished. We list a
few of them, without claim to completeness.
\begin{enumerate}
\item Probabilistic programming languages that incorporate updating
  (conditioning) and/or higher order features, see
  \textit{e.g.}~\cite{DahlqvistDG16,DahlqvistK20,DanosE11,OlmedoGKKM18,StatonYHKW16,DashS20}.

\item The compositional approach to Bayesian
  networks~\cite{CoeckeS12,Fong12} and to Bayesian
  reasoning~\cite{CulbertsonS14,JacobsZ21,Jacobs21c}.

\item The use of diagrammatic methods in (quantum) foundations and
  probability, see~\cite{CoeckeK16} for an overview.

\item Study of `probability monads',
  \textit{e.g.}~in~\cite{Kock12,Jacobs18c}.

\item Axiomatisation of disintegration as key probabilistic technique,
  see \textit{e.g.}~\cite{Fritz20,ChoJ19,FritzR20,FritzGPR20}, and
  also~\cite{ClercDDG17}.

\item Exploration of categorical structures, such as compact closed
  categories~\cite{AbramskyC09,Selinger07} or
  effectuses~\cite{Jacobs15d,ChoJWW15b}.
\end{enumerate}

\noindent These topics cover both ordinary (classical) probability as
well as quantum probability.

An issue that we are particularly interested in is the interplay
between multisets (a.k.a.\ bags) and (probability) distributions (see
\textit{e.g.}~\cite{Jacobs21a}). Multisets play a fundamental role in
probability theory, for instance as representations of urns with
coloured balls, and also of draws from such urns. More generally, in
learning, collections of data items, possibly occurring multiple
times, are properly represented as multisets. Multinomial and
hypergeometric distributions assign probabilities to draws from an
urn, and can thus be represented as distributions on multisets.
Multinomial distributions capture draws \emph{with} replacement,
whereas hypergeometric distributions capture draws \emph{without}
replacement. In the hypergeometric case the number of balls in the urn
decreases with every draw, but in the multinomial case the urn remains
unchanged --- and can thus be represented as a distribution.

A basic, unsolved question that arises is: should one axiomatise
distributions inside the world of multisets (via causal maps, as
\textit{e.g.}~in~\cite[Sec.~6]{ChoJWW15b} or~\cite{CoeckeHK14}), or
should multisets be described in the world of distributions? Briefly:
do multisets or distributions come first? The question is highly
relevant for axiomatisation, since for instance, in the world of
multisets one assumes biproducts $\oplus$, whereas in a world with
distributions coproducts $+$ play a leading role. These differences
can also be expressed in terms of preservation properties of
monads~\cite{Kock12}. Of course, there are many more differences, but
also similarities, such as presence of a monoidal structure $\otimes$
for parallel composition.

Most attention so far has gone to the first approach --- with
multisets first. Recently, the author published a
paper~\cite{Jacobs21b} that details the distributive law $\Mlt\Dst
\Rightarrow \Dst\Mlt$ of multisets over distributions, called the
parallel multinomial law $\pml$. As a result, multisets $\Mlt$ can be
lifted to the Kleisli category $\Kl(\Dst)$ of the distribution
monad. Actually, what turned out to be most relevant, and
well-behaved, is the functor $\Mlt[K]$ that takes $K$-sized multisets
only, for a number $K\in\NNO$. Via the lifted functor $\Mlt[K] \colon
\Kl(\Dst) \rightarrow \Kl(\Dst)$ multisets appear in the world of
distributions --- following the second approach, with distributions
first.

The paper~\cite{Jacobs21b} demonstrated that besides the distributive
law, several other probabilistic operations behave well in the paper's
setting, notably multinomial and hypergeometric distributions, and a
new form of zipping for multisets, called multizip. The aim of the
current paper is to reconstruct many of these results
from~\cite{Jacobs21b}, in an axiomatic setting. Multisets $\Mlt[K]$ of
a fixed size will be defined via a suitable quotient
(see~\cite{Joyal86,AdamekV08}), and many operations are then derived
from the associated universal property, including sums and zips of
multisets and multinomial and hypergeometric maps. Any useful
axiomatisation of probability theory should include at least such
basic distributions.  Interestingly, the distributive law $\pml$ that
plays such a central role in~\cite{Jacobs21b} is completely absent
here. The reason is that $\pml$ is used for lifting $\Mlt[K]$ to
$\Kl(\Dst)$ in~\cite{Jacobs21b}, whereas here our aim is to axiomatise
$\Mlt[K]$ on categories like $\Kl(\Dst)$. The existence of $\pml$
justifies what we do here.

The current paper can be read without knowing about~\cite{Jacobs21b},
but familiarity with that paper does help to understand why certain
choices are made here. Our axiomatisation happens in a monoidal
category with copying and discarding --- like in~\cite{ChoJ19}, called
\emph{Markov category} in~\cite{Fritz20}. Here we shall use Markov
categories with finite colimits, plus distributivity of $\otimes$ over
$+$, and what we call `uniform states'. Due to space constraints we
focus solely on the axiomatisation itself, and not on categories that
possibly satisfy these requirements. Our leading examples are the
Kleisli categories $\Kl(\Dst)$ and $\Kl(\Giry)$ of the distribution
and Giry monads $\Dst$ and $\Giry$, for discrete and continuous
probability. We refer to~\cite{Fritz20} (and also~\cite{DashS21}) for
further instances of Markov categories.

The line of axiomatisation proposed here is a first step, with several
loose ends, and is far from completed. Still this direction is already
of interest in this early stage, because it leads to representation of
practically relevant distributions.  Our approach has a clear discrete
focus so far, centered around multinomial and hypergeometric
distributions, even though it applies in categories for continuous
probability too. But it does not cover typical continuous
distributions like normal, beta, gamma, \textit{etc.}, for which the
approach of~\cite{DashS20} could be useful. Our axiomatisation is
based on multisets, and includes sums and multizips of such multisets,
but not tensors of multisets.  Although tensors of multisets are a
basic operation, they do not seem to fit in the current set up,
because they are not natural w.r.t.\ Kleisli maps, see the discussion
at the end of this paper.

This article is organised as follows. It first introduces Markov
categories with colimits and uses them to define multisets in
Sections~\ref{MarkovSec} and~\ref{MultSec}. The additional
probabilistic requirements, in the form of uniform states are defined
in Section~\ref{UniformSec}. Sections~\ref{ArrFlrnSec},
\ref{DelSubsec} and~\ref{SumZipSec} introduce basic operations on
multisets, such as arrangement and frequentist learning,
draw-and-delete, and sums and zips.  Multinomial and hypergeometric
distributions are then defined in Section~\ref{MulnomHypgeomSec} and
basic properties are proven, such as proper interaction with
frequentist learning, with draw-and-delete, and with multiset zipping.

\section{Markov categories with colimits}\label{MarkovSec}

This section briefly introduces the setting in which we will be
working. A \emph{Markov category} is a symmetric monoidal category in
which:
\begin{itemize}
\item the tensor unit is a final object $1$, so that tensors $\otimes$
  have projections $\pi_{1} \colon X\otimes Y \rightarrow X\otimes 1
  \cong X$; this makes the setting affine,
  see~\cite{Jacobs16a,Jacobs18c};

\item each object $X$ caries a comonoid structure $\delta \colon X
  \rightarrow X\otimes X$ for copying, which is commutative and
  associative and satisfies $\pi_{1} \after \delta = \idmap$, and
  interacts appropriately with the monoidal structure
  $(\otimes,1)$. These copiers can be combined to $n$-ary form
  $\delta[K] \colon X \rightarrow X^{K} = X\otimes \cdots \otimes X$.
\end{itemize}

We should emphasise that these copiers $\delta$ are \emph{not}
natural.  In fact, a map $f$ may be called \emph{deterministic} if it
does commute with copying: $(f\otimes f) \after \delta = \delta \after
f$. It is required that all monoidal isomorphisms are deterministic.

We shall assume that our Markov category has finite colimits, with
several additional requirements.
\begin{itemize}
\item The coprojections (colimit injections) $\kappa_{i} \colon X_{i}
  \rightarrow \mathsl{colim}_{i} X_{i}$ are deterministic, and also
  the mediating map induced by deterministic maps is itself
  deterministic;

\item Tensors $\otimes$ distribute over coproducts $+$. It makes sense
  to require that tensors distribute over all finite colimits --- so
  also over coequalisers --- but we need that requirement at one point
  only, see Theorem~\ref{SumMonoidThm}, and so we explicitly require
  it there.
\end{itemize}

\noindent We shall need one more requirement, namely `uniform states',
which will be introduced in Section~\ref{UniformSec}. We shall think
of maps of the form $1\rightarrow X$ as \emph{distributions}, also
called \emph{states}, over $X$. More generally, maps $Y \rightarrow X$
are then $Y$-indexed distributions, which can be understood as
conditional probabilities $p(X\mid Y)$.

In the remainder of this article we shall work in a fixed Markov
category $\cat{C}$ with finite colimits as described above.

\section{Multisets}\label{MultSec}

The formalisation of multiset in our Markov category $\cat{C}$ is a
key, first step in our axiomatisation. We shall focus on multisets of
a fixed size $K$, that is, on multisets with $K$ elements in total,
including multiplicities. Since multisets can be understood as
sequences where the order does not matter, it makes sense to describe
multisets $\Mlt[K](X)$ over $X$ of size $K$ as quotient $X^{K}
\twoheadrightarrow \Mlt[K](X)$ of the object $X^{K} = X\otimes \cdots
\otimes X$ of sequences of length $K$, see
also~\cite{Joyal86,AdamekV08}. This section only contains the
definition and functoriality. The sum and zip of multisets are
introduced later on, once we have seen uniform states.

We write, as usual, $S_{K}$ for the symmetric group of permutations
$\{1,\ldots,K\} \congrightarrow \{1,\ldots,K\}$. Each permutation
$\sigma\in S_{K}$ translates into a (deterministic) isomorphism
$\underline{\sigma} \colon X^{K} \congrightarrow X^{K}$ via the
monoidal isomorphisms.

\begin{definition}
\label{MultDef}
For each number $K\in\NNO$ and $X\in\cat{C}$ write $\acc[K]$ for the
coequaliser of all (interpreted) permutation maps $\underline{\sigma}
\colon X^{K} \congrightarrow X^{K}$, for $\sigma\in S_{K}$, in:
\[ \xymatrix@C+0pc{
X^{K }\ar@/^2.8ex/[rr]\ar@/_2ex/[rr] & 
   \vdots\rlap{\raisebox{0.25em}{$\;\underline{\sigma}$}} & 
    X^{K} \ar@{->>}[rr]^-{\acc[K]} & & \Mlt[K](X)
} \]

\noindent We call $\Mlt[K](X)$ the object of $K$-sized multisets on
$X$. The map $\acc$ is called accumulator; it turns a list into a
multiset by ignoring orderings. We omit the number $K$ in $\acc[K]$
when it is clear from the context.
\end{definition}

Concretely, in a set-theoretic setting one has: $\acc[5](a,b,a,b,b) =
2\ket{a} + 3\ket{b} \in \Mlt[5](\{a,b\})$. We use a ket notation
$\ket{-}$ for multisets, see~\cite{Jacobs21b} for more (set-theoretic)
details.

We collect some basic facts.

\begin{lemma}
\label{MultLem}
Consider the accumulator map $\acc \colon X^{K} \twoheadrightarrow
\Mlt[K](X)$ from Definition~\ref{MultDef}.
\begin{enumerate}
\item \label{MultLemDet} It is deterministic.

\item \label{MultLemPerms} It satisfies $\acc \after
  \underline{\sigma} = \acc$, for each permutation $\sigma\in S_{K}$.

\item \label{MultLemFun} It is a natural transformation $(-)^{K}
  \Rightarrow \Mlt[K]$, when $\Mlt[K]$ is extended to a functor via:
\[ \vcenter{\xymatrix@C-0.5pc{
X^{K }\ar@/^2.8ex/[rr]\ar@/_2ex/[rr]\ar[d]_{f^{K}} & \vdots & 
    X^{K}\ar@{->>}[rr]^-{\acc[K]_{X}}\ar[d]_{f^{K}} & & 
   \Mlt[K](X)\ar@{-->}[d]^{\Mlt[K](f)}
\\
Y^{K }\ar@/^2.8ex/[rr]\ar@/_2ex/[rr] & \vdots & 
    Y^{K}\ar@{->>}[rr]^-{\acc[K]_{Y}} & & \Mlt[K](Y)
}}
\qquad\mbox{for}\quad
f\colon X \rightarrow Y. \]

\item Precomposition with copying gives a $K$-fold unit map $\acc
  \after \delta[K] \colon X \rightarrow X^{K} \rightarrow \Mlt[K](X)$,
  which is not natural in $X$.
\end{enumerate}
\end{lemma}

\begin{myproof}
\begin{enumerate}
\item The accumulator map $\acc$ is deterministic, as coequaliser of
  deterministic maps, see Section~\ref{MarkovSec}.

\item Since $\acc$ is by construction the coequaliser, we have $\acc
  \after \underline{\sigma} = \acc \after \underline{\tau}$ for all
  permutations $\sigma,\tau \in S_{K}$. This holds in particular when
  we choose $\tau$ to be the identity permutation.

\item This works since $f^{K} \after \underline{\sigma} = \underline{\sigma}
\after f^{K}$ for each permutation $\sigma$.

\item Naturality fails, since only deterministic, not arbitrary, maps
  commute with copier $\delta$. \QED
\end{enumerate}
\end{myproof}

\section{Uniform states}\label{UniformSec}

Let $\cat{C}$ be a Markov category as in Section~\ref{MarkovSec}.  For
each $n\in\NNO$ there is an \emph{interpreted number}
$\underline{n}\in\cat{C}$, namely:
\[ \begin{array}{rcl}
\underline{n}
& \coloneqq &
\underbrace{1 + \cdots + 1}_{n\textrm{ times}}
\qquad \mbox{where $1\in\cat{C}$ is the final object (tensor unit).}
\end{array} \]

\noindent We shall use the sums $+$ in this definition of
$\underline{n}$ up to isomorphisms. Clearly, $\underline{1} = 1$ and
$\underline{0} = 0$, as empty sum. Further, $\underline{n} +
\underline{m} \cong \underline{n+m}$. The codiagonal map $\nabla_{n} =
          [\idmap, \ldots, \idmap] \colon \underline{n} \rightarrow 1$
          is the unique map to $1$.


By distributivity of $\otimes$ over $+$ we get a natural isomorphism
$\underline{n}\otimes X \cong X + \cdots + X$ ($n$ times).  A map of
the form $r\colon 1 \rightarrow \underline{n}$ will be called a
\emph{convex series}, of length $n$. Given another such series
$s\colon 1 \rightarrow \underline{m}$ we write:
\[ \xymatrix@C-0.6pc{
r\bullet s \;\coloneqq\; \Big(1\ar[r]^-{r} & 
   \underline{n} = 1 + \cdots + 1\ar[rr]^-{s + \cdots + s} & &
   \underline{m} + \cdots + \underline{m}\Big).
} \]

\noindent Then, up-to-isomorphism, $r\bullet s = s\bullet r$. This
follows from a Kelly-Laplaza style argument~\cite{KellyL80}:
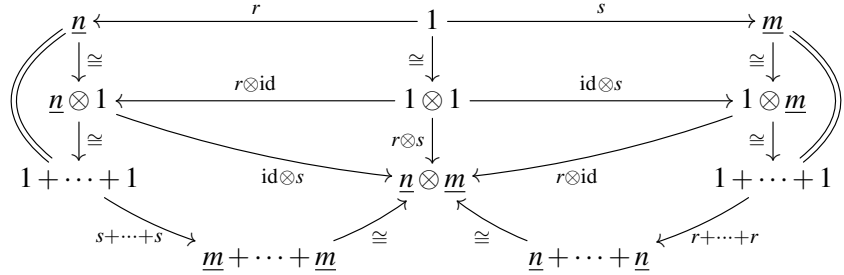
\[ \xymatrix@R-0.8pc@C-0.5pc{
\underline{n}\ar[d]^-{\cong}\ar@{=}@/_5ex/[dd]
   & & 1\ar[ll]_-{r}\ar[rr]^-{s}\ar[d]_{\cong} & & 
   \underline{m}\ar[d]_-{\cong}\ar@{=}@/^5ex/[dd]
\\
\underline{n}\otimes 1\ar[d]^-{\cong}\ar@/_1ex/[drr]_(0.6){\idmap\otimes s}
   & & 1\otimes 1\ar[d]_{r\otimes s}\ar[ll]_-{r\otimes\idmap}
      \ar[rr]^-{\idmap\otimes s} & &
   1\otimes \underline{m}\ar@/^1ex/[dll]^(0.6){r\otimes\idmap}\ar[d]_-{\cong}
\\
1 + \cdots + 1\ar@/_1ex/[dr]_-{s+\cdots+s} & & 
   \underline{n}\otimes\underline{m} & & 
   1 + \cdots + 1\ar@/^1ex/[dl]^-{r+\cdots+r}
\\
& \underline{m} + \cdots + \underline{m}\ar@/_1ex/[ur]_-{\cong} & &
  \underline{n} + \cdots + \underline{n}\ar@/^1ex/[ul]^-{\cong} &
} \]

The following definition is typical for a probabilistic setting.

\begin{definition}
\label{UniformDef}
We say that the category $\cat{C}$ has \emph{uniform states} if for
each $n\geq 1$ there is a \emph{uniform state} $\unif_{n} \colon 1
\rightarrow \underline{n}$. These states are required to satisfy:
\begin{enumerate}
\item $\sigma \after \unif_{n} = \unif_{n}$, for each (interpretated)
  permutation $\sigma\colon \underline{n} \congrightarrow
  \underline{n}$, of size $n$, that is for $\sigma\in S_{n}$;

\item $\unif_{n} \otimes \unif_{m} = \unif_{n\cdot m}$, up-to-isomorphism.
\end{enumerate}
\end{definition}

We think of $\unif \colon 1 \rightarrow \underline{n}$ as the $n$
probabilities $\big(\frac{1}{n}, \ldots, \frac{1}{n}\big)$ adding up to $1$,
and thus forming a convex series. We can use them to form other
convex series, such as:
\[ \xymatrix{
\big(\frac{1}{6}, \frac{1}{2}, \frac{1}{3}\big)
\;\coloneqq\; \Big(1\ar[r]^-{\unif} & 
   \underline{6} = 1 + \underline{3} + 
   \underline{2}\ar[rr]^-{\idmap[1] + \nabla_{3} + \nabla_{2}} & &
   1 + 1 + 1 = \underline{3}\Big).
} \]

\noindent In this way we can form each `fractional' convex series
$\big(\frac{n_1}{n}, \ldots, \frac{n_k}{n}\big)$ as map $1 \rightarrow
\underline{k}$ with $n = \sum_{i}n_i$.
    
Given a convex series $r\colon 1 \rightarrow \underline{n}$, and an
$n$-tuple of maps $f_{i} \colon X \rightarrow Y$ we can form the
\emph{convex sum} $\sum_{i} r\cdot f_{i} \colon X \rightarrow
Y$ via distributivity of $\otimes$ over $+$.
\[ \xymatrix@C-0.3pc{
\sum_{i} r\cdot f_{i}
\;\coloneqq\;\Big(X\cong X\otimes 1 \ar[r]^-{\idmap\otimes r} &
   X\otimes\underline{n} \cong X+\cdots+ X\ar[rr]^-{[f_{1}, \ldots, f_{n}]} & &
   Y\Big).
} \]

\begin{lemma}
\label{ConvexLem}
Consider convex series $r,s$ with suitably typed maps.
\begin{enumerate}
\item \label{ConvexLemSeqComp} Convex sums are preserved by
sequential composition:
\[ \begin{array}{rclcrcl}
\big(\sum_{i} r\cdot f_{i}\big) \after g
& = &
\sum_{i} r\cdot (f_{i} \after g)
& \qquad &
h \after \big(\sum_{i} r\cdot f_{i}\big)
& = &
\sum_{i} r\cdot (h \after f_{i}).
\end{array} \]

\item \label{ConvexLemParComp} Convex sums are preserved by
parallel composition:
\[ \begin{array}{rclcrcl}
\sum_{i} r\cdot (f_{i}\otimes g)
& = &
(\sum_{i} r\cdot f_{i})\otimes g
& \qquad &
\sum_{i} r\cdot (h\otimes f_{i})
& = &
h \otimes (\sum_{i} r\cdot f_{i}).
\end{array} \]

\item \label{ConvexLemConst} Convex sums of constant collections are constant:
\[ \begin{array}{rclcrcl}
\sum_{i} r\cdot f
& = &
f
& \qquad\mbox{and in particular} \qquad &
\sum_{i} r\cdot \idmap
& = &
\idmap.
\end{array} \]

\item \label{ConvexLemComp} $\big(\sum_{j} s\cdot g_{j}\big) \after
  \big(\sum_{i} r\cdot f_{i}\big) = \sum_{j,i} (s\bullet r)\cdot
  (g_{j} \after f_{i})$. \QED
\end{enumerate}
\end{lemma}

\section{Arrangement and frequentist learning}\label{ArrFlrnSec}

In this section we combine multisets with convex sums to obtain
arrangement and frequentist learning operations $\arr \colon
\Mlt[K](X) \rightarrow X^{K}$ and $\flrn \colon \Mlt[K](X) \rightarrow
X$.  Intuitively, the arrangement map $\arr$ sends a multiset to the
uniform distribution of all sequences that accumulate to the
multiset. And the frequentist learning map $\flrn$ normalises a
multiset into a distribution.

Since the symmetric group $S_K$ of permutations of a set with $K$
elements has $K!$ elements we can define:
\begin{equation}
\label{PermEqn}
\begin{array}{rcl}
\perm
& \coloneqq &
\displaystyle\sum_{\sigma\in S_{K}}\,
   \unif_{K!}\cdot\underline{\sigma} \;\colon\; X^{K} \longrightarrow X^{K}.
\end{array}
\end{equation}

The following facts follow readily from Lemma~\ref{ConvexLem}.

\begin{lemma}
\label{PermLem}
\begin{enumerate}
\item \label{PermLemNat} $\perm$ is natural in $X$;

\item \label{PermLemCopy} $\perm \after \delta[K] = \delta[K]$;

\item \label{PermLemNatAcc} $\acc \after \perm = \acc$, for the
  accumulation map of Definition~\ref{MultDef}.
\end{enumerate}
\end{lemma}

\begin{myproof}
The first two points follow from Lemma~\ref{ConvexLem} in:
\[ \begin{array}{rcccccl}
\perm \after f^{K}
& = &
\displaystyle\sum_{\sigma\in S_{K}}\,
   \unif_{K!}\cdot(\underline{\sigma}\after f^{K})
& = &
\displaystyle\sum_{\sigma\in S_{K}}\,
   \unif_{K!}\cdot (f^{K} \after \underline{\sigma})
& = &
f^{K} \after \perm
\end{array} \]
\[ \begin{array}{rcccccl}
\perm \after \delta[K]
& = &
\displaystyle\sum_{\sigma\in S_{K}}\,
   \unif_{K!}\cdot(\underline{\sigma}\after \delta[K])
& = &
\displaystyle\sum_{\sigma\in S_{K}}\,
   \unif_{K!}\cdot \delta[K]
& = &
\delta[K].
\end{array} \]

\noindent For the last point we use Lemma~\ref{MultLem}~\eqref{MultLemPerms}.
\[ \begin{array}{rcccccccl}
\acc \after \perm
& = &
\acc \after \left({\displaystyle\sum}_{\sigma}\,\unif_{K!}\cdot
   \underline{\sigma}\right)
& = &
{\displaystyle\sum}_{\sigma}\,\unif_{K!}\cdot(\acc \after \underline{\sigma})
& = &
{\displaystyle\sum}_{\sigma}\,\unif_{K!}\cdot \acc
& = &
\acc.
\end{array} \eqno{\QEDbox} \]
\end{myproof}

For an object $X\in\cat{C}$ and a number $K\in\NNO$ there are
projections $\pi_{i} \colon X^{K} \rightarrow X$, for $1\leq i\leq
K$. Next, write, for $K\geq 1$,
\begin{equation}
\label{CounitEqn}
\begin{array}{rcl}
\varepsilon[K]
& \coloneqq &
\sum_{i} \unif_{K}\cdot \pi_{i} \;\colon\; X^{K} \longrightarrow X.
\end{array}
\end{equation}

\begin{lemma}
\label{CounitLem}
The map $\varepsilon[K]$ in~\eqref{CounitEqn},
\begin{enumerate}
\item \label{CounitLemNat} is natural in $X$;

\item \label{CounitLemOne} is the identity $X \rightarrow X$ for $K=1$;

\item \label{CounitLemId} satisfies $\varepsilon[K] \after \delta[K] =
  \idmap$, for the $K$-fold copier $\delta[K] \colon X \rightarrow
  X^{K}$. \QED
\end{enumerate}
\end{lemma}

\auxproof{
For $f\colon X \rightarrow Y$,
\[ \begin{array}{rcl}
\varepsilon \after f^{K}
& = &
\big(\sum_{i} \unif_{K}\cdot \pi_{i}\big) \after f^{K}
\\
& = &
\sum_{i} \unif_{K}\cdot (\pi_{i} \after f^{K})
\\
& = &
\sum_{i} \unif_{K}\cdot f \after \pi_{i} 
\\
& = &
f \after \big(\sum_{i} \unif_{K}\cdot \pi_{i}\big)
\\
& = &
f \after \varepsilon.
\end{array} \]

\noindent Next,
\[ \begin{array}{rcl}
\varepsilon \after \delta
& = &
\big(\sum_{i} \unif_{K}\cdot \pi_{i}\big) \after \delta
\\
& = &
\sum_{i} \unif_{K}\cdot (\pi_{i} \after \delta)
\\
& = &
\sum_{i} \unif_{K}\cdot \idmap
\\
& = &
\idmap.
\end{array} \]
}

In a next step we define two basic operations associated with
multisets, namely frequentist learning $\flrn$ and arrangment $\arr$.

\begin{definition}
\label{FlrnArrDef}
For $X\in\cat{C}$, the universal property of the coequaliser yields
frequentist learning maps $\flrn\colon \Mlt[K](X) \rightarrow X$, when
$K\geq 1$, and arrangment maps $\arr \colon \Mlt[K](X) \rightarrow
X^{K}$, for all $K\geq 0$, in situations:
\[ \xymatrix@C-1.2pc@R-0.8pc{
X^{K }\ar@/^2.8ex/[rr]\ar@/_2ex/[rr] & \vdots & 
    X^{K}\ar@{->>}[rr]^-{\acc}\ar@/_2ex/[drr]_{\varepsilon[K]} & & 
   \Mlt[K](X)\ar@{-->}[d]^{\flrn}
& \hspace*{5em} &
X^{K }\ar@/^2.8ex/[rr]\ar@/_2ex/[rr] & \vdots & 
    X^{K}\ar@{->>}[rr]^-{\acc}\ar@/_2ex/[drr]_{\perm} & & 
   \Mlt[K](X)\ar@{-->}[d]^{\arr}
\\
& & & & X
& &
& & & & X^{K}
} \]
\end{definition}

These definitions work since for each permutation $\tau \in S_{K}$ one
has:
\[ \begin{array}{rcccccl}
\varepsilon \after \underline{\tau}
& = &
{\displaystyle\sum}_{i}\,
   \unif_{K}\cdot(\pi_{i} \after \underline{\tau})
& = &
{\displaystyle\sum}_{i}\,
   \unif_{K}\cdot\pi_{i}
& = &
\varepsilon
\end{array}
\qquad
\begin{array}{rcccccl}
\perm \after \underline{\tau}
& = &
{\displaystyle\sum}_{\sigma}\,
   \unif_{K!}\cdot(\underline{\sigma} \after \underline{\tau})
& = &
{\displaystyle\sum}_{\sigma}\,
   \unif_{K!}\cdot\underline{\sigma}
& = &
\perm.
\end{array} \]

\begin{lemma}
\label{FlrnArrLem}
In the above situation,
\begin{enumerate}
\item \label{FlrnArrLemFlrnNat} $\flrn$ is a natural transformation
  $\Mlt[K] \Rightarrow \idmap$;

\item \label{FlrnArrLemArrNat} $\arr$ is a natural transformation
  $\Mlt[K] \Rightarrow (-)^{K}$;

\item \label{FlrnArrLemId} $\acc \after \arr = \idmap$

\item \label{FlrnArrLemPerm} $\underline{\sigma} \after \arr = \arr$,
  for each $\sigma\in S_{K}$.
\end{enumerate}
\end{lemma}

\begin{myproof}
\begin{enumerate}
\item For $f\colon X \rightarrow Y$ we have $\flrn \after \Mlt[K](f) =
  f \after \flrn$, since $\acc$ is epic, using
  Lemma~\ref{CounitLem}~\eqref{CounitLemNat}:
\[ \begin{array}{rcccccccl}
\flrn \after \Mlt[K](f) \after \acc
& = &
\flrn \after \acc \after f^{K}
& = &
\varepsilon \after f^{K}
& = &
f \after \varepsilon 
& = &
f \after \flrn \after \acc.
\end{array} \]

\item Similarly we are done by:
\[ \begin{array}{rcccccccl}
\arr \after \Mlt[K](f) \after \acc
& = &
\arr \after \acc \after f^{K}
& = &
\perm \after f^{K}
& = &
f^{K} \after \perm
& = &
f^{K} \after \arr \after \acc.
\end{array} \]

\item The equation $\acc \after \arr = \idmap$ follows from
Lemma~\ref{PermLem}~\eqref{PermLemNatAcc}:
\[ \begin{array}{rcccccl}
\acc \after \arr \after \acc
& = &
\acc \after \perm
& = &
\acc
& = &
\idmap \after \acc.
\end{array} \]

\item Let $\sigma\in S_{K}$ be given. We get $\underline{\sigma}
\after \arr = \arr$ from:
\[ \begin{array}{rcccccccccl}
\underline{\sigma} \after \arr \after \acc
& = &
\underline{\sigma} \after \perm 
& = &
\displaystyle\sum_{\tau\in S_{K}} \unif_{K!}\cdot 
   (\underline{\sigma} \after \underline{\tau})
& = &
\displaystyle\sum_{\tau\in S_{K}} \unif_{K!}\cdot \underline{\tau}
& = &
\perm
& = &
\arr \after \acc.
\end{array} \eqno{\QEDbox} \]
\end{enumerate}
\end{myproof}

The following points are expected but useful to make explicit.

\begin{lemma}
\label{MultSpecialLem}
\begin{enumerate}
\item \label{MultSpecialLemZero} $\Mlt[0](X)$ is final, so $\Mlt[0](X)
  \cong 1$;

\item \label{MultSpecialLemOne} $\acc[1] \colon X \rightarrow
  \Mlt[1](X)$ is an isomorphism, with $\arr[1]$ as inverse;

\item \label{MultSpecialLemFinal} $\Mlt[K](1)$ is also final;

\item \label{MultSpecialLemInitial} $\Mlt[K](0)$ is final for $K=0$
  and initial for $K>0$.
\end{enumerate}
\end{lemma}

\begin{myproof}
\begin{enumerate}
\item Since $X^{0} = 1$ by definition, we get $\acc[0] \colon 1
  \rightarrow \Mlt[0](X)$, obviously satisfying $! \after \acc[0] =
  \idmap \colon 1 \rightarrow 1$. But then also $\acc[0] \after \;! =
  \idmap$ since $\acc$ is epic and $\acc[0] \after \;! \after \acc[0]
  = \acc[0]$.

\item We have $\flrn \after \acc[1] = \varepsilon[1] = \idmap$ by
  Lemma~\ref{CounitLem}~\eqref{CounitLemOne}. But then also $\acc[1]
  \after \flrn = \idmap$ because $\acc[1]$ is epic. Since also
  $\acc[1] \after \arr[1] = \idmap$ we get $\acc[1]^{-1} = \arr[1] =
  \flrn \colon \Mlt[1](X) \rightarrow X$, by
  Lemma~\ref{FlrnArrLem}~\eqref{FlrnArrLemId}.

\item We already know that $\Mlt[K](1)$ is final for $K=0$, by the
  first point.  For $K>0$ we can use frequentist learning and use that
  $1^{K} = 1$, so $\flrn \after \acc[K] = \idmap \colon 1 \rightarrow
  1$.  But then also $\acc[K] \after \flrn = \idmap$ since $\acc[K]$
  is epic.

\item Note that $0^{K} = 1$ for $K=0$ and $0^{K} = 0$ for $K>0$. That
  $\Mlt[0](0)$ is final follows from the first point. For $K>0$
  frequentist learning $\flrn$ is defined, giving $\flrn \after
  \acc[K] = \idmap \colon 0 \rightarrow 0$. But then also $\acc[K]
  \after \flrn = \idmap$ since $\acc[K]$ is epic. \QED
\end{enumerate}
\end{myproof}

Proofs of the next results are relegated to the appendix.  The
notation $\left(\binom{n}{K}\right) = \binom{n+K-1}{K}$ is the
\emph{multichoose} coefficient. It describes the number of multisets
of size $K$ over an $n$-element set, see
\textit{e.g.}~\cite[II~(5.2)]{Feller70I}. The same result can be
obtained in our abstract setting, in point~\eqref{MltCoprodPropNum}
below.

\begin{proposition}
\label{MltCoprodProp}
\begin{enumerate}
\item \label{MltCoprodPropCoprod} For $K\geq 0$, and objects $X,Y$,
\[ \begin{array}{rcl}
\Mlt[K](X+Y) 
& \cong &
\displaystyle\bigoplus\limits_{0\leq i\leq K}\, 
   \Mlt[i](X) \otimes \Mlt[K\!-\!i](Y).
\end{array} \]

\item \label{MltCoprodPropNum} For a number $n\geq 1$,
\[ \begin{array}{rcccl}
\Mlt[K](\underline{n})
& \cong &
\underline{\displaystyle\left(\!\!\binom{n}{K}\!\!\right)}
& = &
\displaystyle\binom{n+K-1}{K}\cdot 1.
\end{array} \]
\end{enumerate}
\end{proposition}

\section{Uniform deletion}\label{DelSubsec}

When we think of a multiset in $\Mlt[K](X)$ as an urn filled with
$K$-many balls with colours in $X$, we would like to have an operation
for randomly drawing a (single) ball from the urn. We shall describe
this as an operation $\drawdelete \colon \Mlt[K\!+\!1](X) \rightarrow
\Mlt[K](X)$, which we call draw-and-delete.

We fix $K\in\NNO$ and $X\in\cat{C}$. For $1\leq i\leq K+1$ we first
define maps that remove the $i$-th element, and then a uniform
deletion map:
\[ \begin{array}{rclcrcl}
\widehat{\pi}_{i}
& \coloneqq &
\underbrace{\idmap\otimes\cdots\otimes\idmap}_{i-1\mbox{ times}} \otimes\, ! \otimes
   \underbrace{\idmap\otimes\cdots\otimes\idmap}_{K+1-i\mbox{ times}} 
   \;\,\colon\, X^{K+1} \rightarrow X^{K}
& \;\mbox{ and }\; &
\del[K]
& \coloneqq &
\displaystyle\sum_{1\leq i\leq K+1} \unif_{K+1}\cdot \widehat{\pi}_{i}
   \,\colon\, X^{K+1} \rightarrow X^{K}.
\end{array} \]

\noindent In this definition of $\widehat{\pi}_{i}$ we write $!$ for
the map to the final object $1$.

\begin{lemma}
\label{DelLem}
In the above situation,
\begin{enumerate}
\item \label{DelLemPiDelNat} the maps $\widehat{\pi}_{i}$ and $\del$
  are natural;
\item \label{DelLemDiag} deletion commutes with permutation and with
  $\varepsilon$, as in:
\[ \xymatrix@R-0.8pc{
X^{K+1}\ar[d]_{\del}\ar[rr]^-{\perm[K\!+\!1]} & & X^{K+1}\ar[d]^-{\del}
& &
X^{K}\ar@/_1.5ex/[dr]_-{\varepsilon[K]} & & 
   X^{K+1}\ar[ll]_-{\del}\ar@/^1.5ex/[dl]^-{\varepsilon[K+1]}
\\
X^{K}\ar[rr]^-{\perm[K]} & & X^{K}
& &
& X &
} \]

\item \label{DelLemCopy} $\del \after \delta[K\!+\!1] = \delta[K]$;

\item \label{DelLemPi} $\del \after \perm[K\!+\!1] = \pi \after
  \perm[K\!+\!1]$, for the projection $\pi = \idmap\otimes\,! \colon
  X^{K+1} \rightarrow X^{K}$, and then also $\del \after \arr[K\!+\!1]
  = \pi \after \arr[K\!+\!1]$.
\end{enumerate}
\end{lemma}

\begin{myproof}
The first point is obvious, but the other ones requires more
care. We use that for each permutation $\sigma \in S_{K+1}$ and index
$1\leq i\leq K+1$ there is a permutation $\tau\in S_{K}$ and index $j$
with $\widehat{\pi}_{i} \after \underline{\sigma} = \underline{\tau}
\after \widehat{\pi}_{j}$. This yields $K+1$ times the same
$\tau$. Hence:
\[ \begin{array}[b]{rcl}
\del \after \perm[K\!+\!1]
& = &
\displaystyle \left(\sum_{1\leq i\leq K+1} \unif_{K+1}\cdot \widehat{\pi}_{i}\right)
   \after \left(\sum_{\sigma\in S_{K+1}} \unif_{(K+1)!}\cdot \underline{\sigma}\right)
\\[+1.5em]
& = &
\displaystyle \sum_{1\leq i\leq K+1}\sum_{\sigma\in S_{K+1}} 
   (\unif_{K+1}\otimes\unif_{(K+1)!})\cdot 
   (\widehat{\pi}_{i}\otimes \underline{\sigma})
\\[+1.2em]
& = &
\displaystyle \sum_{1\leq i\leq K+1}\sum_{\tau\in S_{K}} 
   (\unif_{K+1}\otimes\unif_{K!}\otimes\unif_{K+1})\cdot 
   (\widehat{\pi}_{i}\otimes \underline{\tau} \otimes\idmap)
\\[+1.2em]
& = &
\displaystyle \sum_{1\leq i\leq K+1}\sum_{\tau\in S_{K}} 
   (\unif_{K+1}\otimes\unif_{K!})\cdot 
   (\widehat{\pi}_{i}\otimes \underline{\tau})
\hspace*{\arraycolsep}=\hspace*{\arraycolsep}
\perm[K] \after \del.
\end{array} \]

\noindent Similarly, all composites $\pi_{j} \after \widehat{\pi}_{i}$
consist of $K$ times the projection $\pi_{i} \colon X^{K+1} \rightarrow X$.
Hence:
\[ \begin{array}[b]{rcl}
\varepsilon[K] \after \del
& = &
\displaystyle \left(\sum_{1\leq j\leq K} \unif_{K}\cdot \pi_{j}\right)
   \after \left(\sum_{1\leq i\leq K+1} \unif_{K+1}\cdot \widehat{\pi}_{i}\right)
\\[+1.5em]
& = &
\displaystyle \sum_{1\leq j\leq K} \sum_{1\leq i\leq K+1}
   (\unif_{K} \otimes \unif_{K+1}) \cdot (\idmap\otimes\pi_{i})
\hspace*{\arraycolsep}=\hspace*{\arraycolsep}
\displaystyle \sum_{1\leq i\leq K+1} \unif_{K+1} \cdot \pi_{i}
\hspace*{\arraycolsep}=\hspace*{\arraycolsep}
\varepsilon[K\!+\!1].
\end{array} \]

\noindent Along the same lines we obtain point~\eqref{DelLemPi}.
\[ \begin{array}[b]{rcccccl}
\pi \after \perm[K\!+\!1]
& = &
\displaystyle\sum_{\sigma\in S_{K+1}} \unif_{(K+1)!} \cdot 
   (\pi \after \underline{\sigma})
& = &
\displaystyle\sum_{\tau\in S_{K}} \sum_{1\leq i\leq K+1} 
   (\unif_{K!} \otimes \unif_{K+1}) \cdot 
   (\underline{\tau}\otimes\pi_{i})
& = &
\perm[K] \after \del.
\end{array} \]

\noindent Finally, for point~\eqref{DelLemCopy} we use:
\[ \begin{array}[b]{rcccccccl}
\del \after \delta[K\!+\!1]
& = &
\displaystyle\sum_{1\leq i\leq K+1} \unif_{K+1}\cdot 
   (\widehat{\pi}_{i} \after \delta[K\!+\!1])
& = &
\displaystyle\sum_{1\leq i\leq K+1} \unif_{K+1}\cdot \delta[K]
& = &
\delta[K].
\end{array} \eqno{\QEDbox} \]
\end{myproof}

These results allow us to define a draw-and-delete map
$\drawdelete\colon \Mlt[K\!+\!1](X) \rightarrow \Mlt[K](X)$ in:
\begin{equation}
\label{DrawDeleteEqn}
\vcenter{\xymatrix@C-1.2pc@R-1.5pc{
X^{K+1}\ar@/^2.8ex/[rr]\ar@/_2ex/[rr] & \vdots & 
    X^{K+1}\ar@{->>}[rr]^-{\acc}\ar@/_1ex/[dr]_-{\del} & & 
   \Mlt[K\!+\!1](X)\ar@{-->}[dd]^-{\drawdelete}
\\
& & & X^{K}\ar@/_1ex/[dr]_-{\acc} &
\\
& & & & \Mlt[K](X)
}}
\end{equation}

\begin{proposition}
\label{DrawDeleteProp}
Consider the draw-and-delete map $\drawdelete$ defined
in~\eqref{DrawDeleteEqn}.  Frequentist learning after draw-and-delete
is frequentist learning, as described on the left below.
\[ \xymatrix@C-0.5pc@R-0.8pc{
\Mlt[K](X)\ar@/_1.5ex/[dr]_-{\flrn} & & 
   \Mlt[K\!+\!1](X)\ar[ll]_-{\drawdelete}\ar@/^1.5ex/[dl]^-{\flrn}
& \hspace*{3em} &
X^{K+1}\ar[d]_-{\del}\ar@{->>}[r]^-{\acc} &
   \Mlt[K\!+\!1](X)\ar[r]^-{\arr}\ar[d]_-{\drawdelete} & X^{K+1}\ar[d]^-{\del}
\\
& X &
& &
X^{K}\ar@{->>}[r]^-{\acc} & \Mlt[K](X)\ar[r]^-{\arr} & X^{K}
} \]

\noindent In addition, the rectangles on the right commute.
\end{proposition}

\begin{myproof}
The above triangle is obtained via the commuting triangle in
Lemma~\ref{DelLem}:
\[ \begin{array}{rcccccccl}
\flrn \after \drawdelete \after \acc
& = &
\flrn \after \acc \after \del
& = &
\varepsilon \after \del
& = &
\varepsilon
& = &
\flrn \after \acc.
\end{array} \]

\noindent The outer rectangle on the right commutes since it is the
rectangle in Lemma~\ref{DelLem}. The inner rectangle on the left
commutes by definition~\eqref{DrawDeleteEqn} of draw-and-delete.
Hence the inner rectangle on the right commutes because $\acc$ is
epic. \QED
\end{myproof}

\section{Sum and zip of multisets}\label{SumZipSec}

This section introduces two binary operations on multisets, namely the
sum and zip. The sum is well-known and involves addition of
multplicities. The zip of multisets is a recently introduced operation
(in~\cite{Jacobs21b}) that is more complicated. It will be called
multizip, to distinguish it from the zip operation for lists. Both
operations are obtained basically in the same way, namely by:
(1)~turning multisets into lists, via arrangement; (2)~performing the
corresponding operation on lists; (3)~turning the result back into a
multiset via accumulation.

\subsection{Summing multisets}\label{SumSubsec}

Concatenation $\concat$ of lists of fixed lengths can be described
in a monoidal category as deterministic map of the form:
\[ \xymatrix{
X^{K}\otimes X^{L}\ar[r]^-{\concat}_-{\cong} & X^{K+L}.
} \]

\noindent We use it in the following way to define a sum of
multisets.

\begin{definition}
\label{SumDef}
For $K\in\NNO$ and $X\in\cat{C}$ define $+\colon
\Mlt[K](X)\otimes\Mlt[L](X) \rightarrow \Mlt[K\!+\!L](X)$ as
composite:
\[ \xymatrix@C-0.3pc{
+ \;\coloneqq\; \Big(\Mlt[K](X)\otimes\Mlt[L](X)\ar[rr]^-{\arr\otimes\arr}
   & & X^{K}\otimes X^{L}\ar[r]^-{\concat}_-{\cong} & 
   X^{K+L}\ar[r]^-{\acc} & \Mlt[K\!+\!L](X)\Big).
} \]
\end{definition}

Since $\arr$ and $\acc$ are natural, and obviously concatenation
$\concat$ too, so the composite defining $+$ in
Definition~\ref{SumDef} is natural too.

\begin{lemma}
\label{SumMonoidLem}
The sum $+$ of multisets from Definition~\ref{SumDef} is
commutative and associative, satisfying:
\[ \xymatrix@R-1pc{
\big(\Mlt[K](X)\otimes\Mlt[L](X)\big)\otimes\Mlt[N](X)
   \ar[r]^-{+\otimes\idmap}\ar[dd]_{\alpha}^-{\cong}
   & \Mlt[K\!+\!L](X)\otimes\Mlt[N](X)\ar[d]^-{+}
\\
& \Mlt[K\!+\!L\!+\!N](X)
\\
\Mlt[K](X)\otimes\big(\Mlt[L](X)\otimes\Mlt[N](X)\big)
   \ar[r]^-{\idmap\otimes +}
   & \Mlt[K](X)\otimes\Mlt[L\!+\!N](X)\ar[u]_-{+}
} \]
\[ \xymatrix@R-0.8pc@C-1pc{
\Mlt[K](X)\otimes\Mlt[L](X)\ar[d]_{\gamma}^-{\cong}\ar[r]^-{+} 
   & \Mlt[K\!+\!L](X)\ar@{=}[d]
& \hspace*{3em} &
\Mlt[K](X)\ar[r]^-{\cong}\ar@{=}@/_2ex/[dr] 
   & \Mlt[0](X)\otimes\Mlt[K](X)\ar[d]^{+}
\\
\Mlt[L](X)\otimes\Mlt[K](X)\ar[r]^-{+} & \Mlt[L\!+\!K](X)
& &
& \Mlt[K](X)
} \]
\end{lemma}

Via this associativity and commutativity of $+$ we can define an
$K$-fold sum, for $n\geq 1$,
\begin{equation}
\label{MultSumDiag}
\vcenter{\xymatrix{
\Mlt[L](X)^{K}\ar[r]^-{\sum_K} & \Mlt[K\cdot L](X)
& \mbox{and then also} &
\Mlt[K]\big(\Mlt[L](X))\ar[r]^-{\mu_{K,L}} & \Mlt[K\cdot L](X).
}}
\end{equation}

\begin{theorem}
\label{SumMonoidThm}
Assume that maps of the form $\acc\otimes\acc$ are coequaliser too,
\textit{e.g.}~because $\otimes$ preserves coequalisers.
\begin{enumerate}
\item The sum of multisets $+$ from Lemma~\ref{SumMonoidLem} satisfies
  $+ \after (\acc\otimes\acc) = \acc \after \concat$, and is thus a
  (mediating) deterministic map.

\item The maps $\sum_{K}$ and $\mu_{K,L}$ in~\eqref{MultSumDiag} are
  natural.

\item The maps $\mu_{K,L}$ in~\eqref{MultSumDiag}, together with the
  maps $\acc[1] \colon X \rightarrow \Mlt[1](X)$ from
  Lemma~\ref{MultSpecialLem}~\eqref{MultSpecialLemOne}, turn $\Mlt[K]$
  into a graded monad, see \textit{e.g.}~\cite{MiliusPS15,FujiiKM16},
  with respect to the multiplicative monoid $(\NNO, \cdot, 1)$ of
  natural numbers.
\end{enumerate}
\end{theorem}

\begin{myproof}
\auxproof{
\[ \begin{array}{rcl}
+ \after (\acc\otimes\acc)
& = &
\acc \after \concat \after (\perm\otimes\perm)
\\
& = &
\displaystyle\sum_{\sigma\in S_{K}, \tau\in S_{L}} (\unif_{K!}\cdot \unif_{L!})
   \cdot \big(\acc \after \concat \after 
   (\underline{\sigma}\otimes\underline{\tau})\big)
\\
& = &
\displaystyle\sum_{\sigma\in S_{K}, \tau\in S_{L}} (\unif_{K!}\cdot \unif_{L!})
   \cdot \big(\acc \after [\underline{\sigma},\underline{\tau}]
   \after \concat\big)
\\
& = &
\displaystyle\sum_{\sigma\in S_{K}, \tau\in S_{L}} (\unif_{K!}\cdot \unif_{L!})
   \cdot \big(\acc \after \concat\big)
\\
& = &
\acc \after \concat.
\end{array} \]
}

The equation in the first point is easy. It makes $+$ deterministic,
as a mediating map for a deterministic map $\acc \after \concat$. The
second point is obtained by using that the sum $\sum_K$ and
multiplications $\mu_{K,L}$ maps in~\eqref{MultSumDiag} are determined
by:
\[ \vcenter{\xymatrix@C+0.5pc{
\big(X^{L}\big)^{K}\ar@{->>}[r]^-{\acc[L]^{K}}\ar[d]_{\concat_K}^{\cong} & 
   \Mlt[L](X)^{K}\ar[r]^-{\sum_K} & \Mlt[K\cdot L](X)
&
\Mlt[L](X)^{K}\ar@{->>}[r]^-{\acc[K]}\ar@/_2ex/[dr]_{\sum_K} & 
   \Mlt[K]\big(\Mlt[L](X))\ar[d]^-{\mu_{K,L}}
\\
X^{K\cdot L}\ar@{->>}@/_2ex/[urr]_{\acc[K\cdot L]} & &
&
& \Mlt[K\cdot L](X)
}} \eqno{\raisebox{-2em}{$\QEDbox$}} \]

\auxproof{
For $f\colon X \rightarrow Y$ we get $\Mlt[K\cdot L](f) \after \sum_{K}
= \sum_{K} \after \Mlt[L](f)^{K}$ in:
\[ \xymatrix@R-0.5pc@C+1pc{
X^{K\cdot L}\ar[ddd]_{f^{K\cdot L}}\ar@/^3ex/[drrr]^-{\acc}
\\
& (X^{L})^{K}\ar[ul]_-{\concat_K}^-{\cong}\ar@{->>}[r]^-{\acc[L]^{K}}
   \ar[d]_{(f^{L})^{K}} &
   \Mlt[L](X)^{K}\ar[r]^-{\sum_K}\ar[d]^{\Mlt[L](f)^{K}} & 
   \Mlt[K\cdot L](X)\ar[d]^{\Mlt[K\cdot L](f)}
\\
& (Y^{L})^{K}\ar[dl]_-{\concat_K}^-{\cong}\ar@{->>}[r]^-{\acc[L]^{K}} &
   \Mlt[L](Y)^{K}\ar[r]^-{\sum_K} & 
   \Mlt[K\cdot L](Y)
\\
Y^{K\cdot L}\ar@/_3ex/[urrr]_-{\acc}
} \]

\noindent This naturality of $\sum_K$ gives naturality of $\mu_{K,L}$ in:
\[ \xymatrix@R-0.5pc@C+1pc{
\Mlt[L](X)^{K}\ar@{->>}[dr]^-{\acc}\ar[ddd]_{\Mlt[L](f)^{K}}\ar@/^3ex/[drrr]^-{\sum_K}
\\
& \Mlt[K]\Mlt[L](X)\ar[rr]^-{\mu_{K,L}}\ar[d]^-{\Mlt[K]\Mlt[L](f)} & & 
   \Mlt[K\cdot L](X)\ar[d]^-{\Mlt[K\cdot L](f)}
\\
& \Mlt[K]\Mlt[L](Y)\ar[rr]^-{\mu_{K,L}} & & \Mlt[K\cdot L](Y)
\\
\Mlt[L](U)^{K}\ar@{->>}[ur]_-{\acc}\ar@/_3ex/[urrr]_-{\sum_K}
} \]
}

\auxproof{
\noindent We need to prove commutation of the following unit laws.
\[ \xymatrix@C+2pc{
\Mlt[K](X)\ar[r]^-{\acc[1]}\ar@/_2ex/@{=}[dr] & 
   \Mlt[1]\Mlt[K](X)\ar[d]^-{\mu_{1,K}}
&
\Mlt[K](X)\ar[r]^-{\Mlt[K](\acc[1])}\ar@/_2ex/@{=}[dr] & 
   \Mlt[K]\Mlt[1](X)\ar[d]^-{\mu_{K,1}}
\\
& \Mlt[K](X)
&
& \Mlt[K](X)
} \]

\noindent The equation on the left is easy, since $\mu_{1,K} \after \acc[1]
= \sum_{1} = \idmap$. Next,
\[ \begin{array}{rcl}
\mu_{K,1} \after \Mlt[K](\acc[1]) \after \acc[K]
& = &
\mu_{K,1} \after \acc[K] \after \acc[1]^{K}
\\
& = &
\sum_{K} \after \acc[1]^{K}
\\
& = &
\acc[1\cdot K] \after \concat_{1}
\\
& = &
\idmap \after \acc[K] 
\end{array} \]

For the associativity diagram we have to do much more work:
\[ \xymatrix{
\Mlt[K]\Mlt[L]\Mlt[N](X)\ar[rr]^-{\mu_{K,L}}\ar[d]_{\Mlt[K](\mu_{L,N})} & & 
   \Mlt[K\cdot L]\Mlt[N](X)\ar[d]^-{\mu_{K\cdot L, N}}
\\
\Mlt[K]\Mlt[L\cdot N](X)\ar[rr]^-{\mu_{K,L\cdot N}} & & \Mlt[K\cdot L\cdot N](X)
} \]

\noindent We precompose both composites in this rectangle with the
horizontal composite in:
\[ \xymatrix@C-0.5pc{
& (X^{N})^{K\cdot L}\ar@/_2ex/[dl]_-{\concat_{K\cdot L}}
\\
X^{K\cdot L\cdot N} & 
   \big((X^{N})^{L}\big)^{K}\ar@{->>}[rr]^-{\big(\acc[N]^{L}\big)^{K}}
   \ar[u]_{\cong}^{\concat_K}\ar[d]^{\cong}_{(\concat_{L})^{K}} & &
    \big(\Mlt[N](X)^{L}\big)^{K}\ar@{->>}[rr]^-{\acc[L]^{K}} & &
    \Mlt[L]\Mlt[N](X)^{K}\ar@{->>}[r]^-{\acc[K]} & \Mlt[K]\Mlt[L]\Mlt[N](X)
\\
& \big(X^{L\cdot N}\big)^{K}\ar@/^2ex/[ul]^-{\concat_{K}}
} \]

\[ \begin{array}{rcl}
\mu_{K\cdot L, N} \after \mu_{K,L} \after \acc[K] \after
   \acc[L]^{K} \after (\acc[N]^{L})^{K}
& = &
\mu_{K\cdot L, N} \after \sum_{K} \after \acc[L]^{K} \after (\acc[N]^{L})^{K}
\\
& = &
\mu_{K\cdot L, N} \after \acc[K\cdot L] \after \concat_{K} \after (\acc[N]^{L})^{K}
\\
& = &
\sum_{K\cdot L} \after \acc[N]^{K\cdot L} \after \concat_{K}
\\
& = &
\sum_{K\cdot L} \after \acc[K\cdot L] \after \concat_{K}
\\
& = &
\acc[K\cdot L\cdot N] \after \concat_{K\cdot L} \after \concat_{K}
\\
& = &
\acc[K\cdot L\cdot N] \after \concat_{K} \after (\concat_{L})^{K}
\\
& = &
\sum_{K} \after (\acc[L\cdot N])^{K} \after (\concat_{L})^{K}
\\
& = &
\sum_{K} \after (\sum_{L})^{K} \after (\acc[N]^{L})^{K}
\\
& = &
\mu_{K, L\cdot N} \after \acc[K] \after (\mu_{L,N})^{K} \after
   \acc[L]^{K} \after (\acc[N]^{L})^{K}
\\
& = &
\mu_{K, L\cdot N} \after \Mlt[K](\mu_{L,N}) \after \acc[K] \after
   \acc[L]^{K} \after (\acc[N]^{L})^{K}.
\end{array} \]
}
\end{myproof}

When $(U,m,u)$ is an internal commutative monoid we can define
composition maps $U^{K} \rightarrow U$ and $m[K]\colon \Mlt[K](U)
\rightarrow U$. The latter commutes with the sum in
Lemma~\ref{SumMonoidLem}: $m \after (m[K]\otimes m[L]) = m[K\!+\!L]
\after +$.

\auxproof{
\[ \begin{array}{rcl}
m[K\!+\!L] \after + \after (\acc\otimes\acc)
& = &
m[K\!+\!L] \after \acc \after \concat 
\\
& = &
m_{K+L} \after \concat
\\
& = &
m \after (m_{K}\otimes m_{L})
\\
& = &
m \after (m[K]\otimes m[L]) \after (\acc\otimes\acc)
\end{array} \]
}

\subsection{Zipping multisets}\label{MultizipSubsec}

In functional programming there is the familiar zip operation
$X^{K}\times Y^{K} \congrightarrow (X\times Y)^{K}$ that pairs the
items of two lists of the same length. It also exists in a monoidal
category, via rearrangement:
\[ \xymatrix@R-1.2pc@C-1.0pc{
\zip \;\coloneqq\; \Big(X^{K}\otimes Y^{K}\ar@{=}[r] & 
   \big(X\otimes\cdots\otimes X\big)\otimes\big(Y\otimes\cdots\otimes Y\big)
   \ar[r]^-{\cong}
& \big(X\otimes Y\big) \otimes \cdots \otimes \big(X\otimes Y\big)
   \ar@{=}[r] & \big(X\otimes Y\big)^{K}\Big).
} \]

\noindent Clearly, this $\zip$ is natural in $X,Y$. We can now define
an analogous zip operation for multisets of the same size, called
\emph{multizip}, and written as $\mzip$.  It makes the multiset
functor $\Mlt[K]$ monoidal.

\begin{definition}
\label{MzipDef}
For $K\in\NNO$ and $X,Y\in\cat{C}$ define $\mzip\colon
\Mlt[K](X)\otimes\Mlt[K](Y) \rightarrow \Mlt[K](X\times Y)$ as
composite:
\[ \xymatrix@C-0.5pc{
\mzip \;\coloneqq\; \Big(\Mlt[K](X)\otimes\Mlt[K](Y)\ar[rr]^-{\arr\otimes\arr}
   & & X^{K}\otimes Y^{K}\ar[r]^-{\zip}_-{\cong} & 
   (X\otimes Y)^{K}\ar[r]^-{\acc} & \Mlt[K](X\otimes Y)\Big).
} \]
\end{definition}

\begin{proposition}
\label{MzipProp}
\begin{enumerate}
\item \label{MzipPropNat} Multizip is natural.

\item \label{MzipPropArr} Arrangement commutes with $\zip$ and
  $\mzip$, as in:
\[ \xymatrix@R-0.8pc{
\Mlt[K](X)\otimes\Mlt[K](Y)\ar[d]_{\mzip}\ar[rr]^-{\arr\otimes\arr} & &
   X^{K}\otimes Y^{K}\ar[d]_-{\cong}^-{\zip}
\\
\Mlt[K](X\otimes Y)\ar[rr]^-{\arr} & & (X\otimes Y)^{K}
} \]

\item Multizip is associative, making $\Mlt[K]$ together with the
  isomorphism $1 \congrightarrow \Mlt[K](1)$ from
  Lemma~\ref{MultSpecialLem}~\eqref{MultSpecialLemOne} a monoidal
  functor.

\item Multizip commutes with projections: $\Mlt[K](\pi_{1}) \after
  \mzip = \pi_{1} \colon \Mlt[K](X)\otimes\Mlt[K](Y) \rightarrow
  \Mlt[K](X)$, and similarly for the second projection $\pi_{2}$.

\item \label{MzipPropDD} Multizip commutes with draw-and-delete:
\[ \xymatrix@R-0.8pc{
\Mlt[K\!+\!1](X)\otimes\Mlt[K\!+\!1](Y)\ar[d]_{\mzip}
   \ar[rr]^-{\drawdelete\otimes\drawdelete} & & 
   \Mlt[K](X)\otimes\Mlt[K](Y)\ar[d]^{\mzip}
\\
\Mlt[K\!+\!1](X\otimes Y)\ar[rr]^-{\drawdelete} & & \Mlt[K](X\otimes Y)
} \]
\end{enumerate}
\end{proposition}

\begin{myproof}
\begin{enumerate}
\item Easy, since all ingredients in the definition of $\mzip$
are natural.

\item Since:
\[ \begin{array}{rcl}
\arr \after \mzip
\hspace*{\arraycolsep}=\hspace*{\arraycolsep}
\perm \after \zip \after (\arr\otimes\arr)
& = &
\displaystyle\sum_{\sigma\in S_{K}} \unif_{K!}\cdot 
   \big(\underline{\sigma} \after \zip \after (\arr\otimes\arr)\big)
\\[+1.2em]
& = &
\displaystyle\sum_{\sigma\in S_{K}} \unif_{K!}\cdot \big(\zip \after
   (\underline{\sigma}\otimes\underline{\sigma}) \after (\arr\otimes\arr)\big)
\\[+1.2em]
& = &
\displaystyle\sum_{\sigma\in S_{K}} \unif_{K!}\cdot 
   \big(\zip \after (\arr\otimes\arr)\big)
   \qquad \mbox{by Lemma~\ref{FlrnArrLem}~\eqref{FlrnArrLemPerm}}
\\[+0.5em]
& = &
\zip \after (\arr\otimes\arr).
\end{array} \]

\item We reason as follows, using associativity of $\zip$, and
  ignoring monoidal associativity.
\[ \begin{array}{rcll}
\mzip \after (\mzip\otimes\idmap)
& = &
\acc \after \zip \after ((\arr\after\mzip)\otimes\arr)
\\
& = &
\acc \after \zip \after (\zip\otimes\idmap) \after
   (\arr\otimes\arr\otimes\arr) \quad
   & \mbox{by point~\eqref{MzipPropArr}}
\\
& = &
\acc \after \zip \after (\idmap\otimes\zip) \after
   (\arr\otimes\arr\otimes\arr)
   & \mbox{by associativity of $\zip$}
\\
& = &
\acc \after \zip \after (\arr\otimes(\arr\after\mzip))
   & \mbox{by point~\eqref{MzipPropArr} again}
\\
& = &
\mzip \after (\idmap\otimes\mzip).
\end{array} \]

\item We do the computation for the first projection $\pi_{1}\colon
X\otimes Y \rightarrow X$.
\[ \begin{array}{rcl}
\Mlt[K](\pi_{1}) \after \mzip
\hspace*{\arraycolsep}=\hspace*{\arraycolsep}
\Mlt[K](\pi_{1}) \after \acc \after \zip \after (\arr\otimes\arr)
& = &
\acc \after (\pi_{1})^{K} \after \zip \after (\arr\otimes\arr)
\\
& = &
\acc \after \pi_{1} \after (\arr\otimes\arr)
\\
& = &
\acc \after \arr \after \pi_{1}
\\
& = &
\pi_{1} \qquad \mbox{by Lemma~\ref{FlrnArrLem}~\eqref{FlrnArrLemId}.}
\end{array} \]

\item Via the following argument:
\[ \begin{array}[b]{rcll}
\mzip \after (\drawdelete\otimes\drawdelete)
& = &
\acc \after \zip \after 
   \big((\arr \after \drawdelete)\otimes(\arr \after \drawdelete)\big) \quad
\\
& = &
\acc \after \zip \after 
   \big((\del \after \arr)\otimes(\del \after \arr)\big)
   & \mbox{by Proposition~\ref{DrawDeleteProp}}
\\
& = &
\acc \after \zip \after 
   \big((\pi \after \arr)\otimes(\pi \after \arr)\big)
   & \mbox{by Lemma~\ref{DelLem}~\eqref{DelLemPi}}
\\
& = &
\acc \after \pi \after \zip \after (\arr\otimes\arr)
\\
& = &
\acc \after \pi \after \arr \after \mzip
   & \mbox{by point~\eqref{MzipPropArr}}
\\
& = &
\acc \after \del \after \arr \after \acc \after \zip \after (\arr\otimes\arr)
   & \mbox{by Lemma~\ref{DelLem}~\eqref{DelLemPi}}
\\
& = &
\drawdelete \after \acc \after \arr \after \acc \after \zip \after 
   (\arr\otimes\arr)
\\
& = &
\drawdelete \after \acc \after \zip \after (\arr\otimes\arr)
\\
& = &
\drawdelete \after \mzip.
\end{array} \eqno{\QEDbox} \]
\end{enumerate}
\end{myproof}

\section{Multinomial and hypergeometric distributions}\label{MulnomHypgeomSec}

This section finally introduces multinomial and hypergeometric
distributions in the current axiomatic setting. The ensuing results
are as in~\cite{Jacobs21b} for the Kleisli category $\Kl(\Dst)$ of the
distribution monad, but are now obtained in a general categorical
setting.

\begin{definition}
\label{MulnomHypgeomDef}
\begin{enumerate}
\item For an arbitrary map $f\colon X \rightarrow Y$ and number
  $K\in\NNO$ we define the $K$-sized multinomial $\multinomial[K](f)
  \colon X \rightarrow \Mlt[K](Y)$ of $f$ as:
\[ \xymatrix{
\multinomial[K](f) \;\coloneqq\; \Big(X\ar[r]^-{\delta[K]} &
   X^{K}\ar[r]^-{f^{K}} & Y^{K}\ar[r]^-{\acc} & \Mlt[K](Y)\Big).
} \]

\item For $L\geq K$ define the hypergeometric map $\hypergeometric[L,K]
\colon \Mlt[L](X) \rightarrow \Mlt[K](X)$ via repeated draw-and-delete:
\[ \begin{array}{rcl}
\hypergeometric[L,K]
& \coloneqq &
\underbrace{\drawdelete \after \cdots \after \drawdelete}_{L-K\text{ times}}
   \;\colon\; \Mlt[L](X) \longrightarrow \Mlt[K](X).
\end{array} \]
\end{enumerate}
\end{definition}

We first prove several results about multinomials.

\begin{theorem}
\label{MulnomThm}
The multinomial maps satisfy the following properties.
\[ \xymatrix@R-0.8pc@C+1.5pc{
X\ar[r]^-{\multinomial[K](f)}\ar[d]_{f} & \Mlt[K](Y)\ar[d]^-{\arr}
&
X\ar[r]^-{\multinomial[K](f)}\ar@/_2ex/[dr]_-{f} & \Mlt[K](Y)\ar[d]^-{\flrn}
\\
Y\ar[r]^-{\delta[K]} & Y^{K}
&
& Y
} \]
\[ \xymatrix@R-0.8pc@C+3pc{
X\ar[r]^-{\multinomial[K\!+\!1](f)}\ar@/_2ex/[dr]_-{\multinomial[K](f)\quad} & 
   \Mlt[K\!+\!1](Y)\ar[d]^-{\drawdelete}
&
X\ar[r]^-{\multinomial[K](\multinomial[L](f))}
   \ar@/_2ex/[dr]_-{\multinomial[K\cdot L](f)\quad} & 
   \Mlt[K]\big(\Mlt[L](Y))\ar[d]^{\mu_{K,L}}
\\
& \Mlt[K](Y)
&
& \Mlt[K\cdot L](Y)
} \]
\[ \hspace*{-1em}\xymatrix@R-0.8pc@C-0.6pc{
X\ar[r]^-{\delta}\ar@/_2ex/[drrrr]_-{\multinomial[K\!+\!L](f)\quad} & 
   X\otimes X\ar[rrr]^-{\;\;\multinomial[K](f)\otimes\multinomial[L](f)}
   & & & \Mlt[K](Y)\otimes\Mlt[L](Y)\ar[d]^-{+}
&
X\otimes A\ar[rrr]^-{\;\multinomial[K](f)\otimes\multinomial[K](g)}
   \ar@/_2ex/[drrr]_-{\multinomial[K](f\otimes g)\quad}
   & & & \Mlt[K](Y)\otimes\Mlt[K](B)\ar[d]^-{\mzip}
\\
& & & & \Mlt[K\!+\!L](Y)
&
& & & \Mlt[K](Y\otimes B)
} \]
\end{theorem}






\begin{myproof}
We handle commutation of the six digrams one by one. The first one
follows from Lemma~\ref{PermLem}:
\[ \begin{array}{rcccccccl}
\arr \after \multinomial[K](f)
& = &
\arr \after \acc \after f^{K} \after \delta[K]
& = &
\perm \after f^{K} \after \delta[K]
& = &
f^{K} \after \perm \after \delta[K]
& = &
f^{K} \after \delta[K].
\end{array} \]

\noindent Via Lemma~\ref{CounitLem}:
\[ \begin{array}{rcccccccl}
\flrn \after \multinomial[K](f)
& = &
\flrn \after \acc \after f^{K} \after \delta[K]
& = &
\varepsilon[K] \after f^{K} \after \delta[K]
& = &
f \after \varepsilon[K] \after \delta[K]
& = &
f.
\end{array} \]

\noindent Next, by Lemma~\ref{DelLem}~\eqref{DelLemPiDelNat}
and~\eqref{DelLemCopy},
\[ \begin{array}[b]{rcl}
\drawdelete \after \multinomial[K\!+\!1](f)
& = &
\drawdelete \after \acc \after f^{K+1} \after \delta[K\!+\!1]
\\
& = &
\acc \after \del \after f^{K+1} \after \delta[K\!+\!1]
\\
& = &
\acc \after f^{K} \after \del \after \delta[K\!+\!1]
\hspace*{\arraycolsep}=\hspace*{\arraycolsep}
\acc \after f^{K} \after \delta[K]
\hspace*{\arraycolsep}=\hspace*{\arraycolsep}
\multinomial[K](f).
\end{array} \]

\noindent Next, we use the diagrams from the proof of 
Theorem~\ref{SumMonoidThm}.
\[ \begin{array}[b]{rcl}
\mu_{K,L} \after \multinomial[K]\big(\multinomial[L](f)\big)
& = &
\mu_{K,L} \after \acc[K] \after \multinomial[L](f)^{K} \after \delta[K]
\\
& = &
\sum_{K} \after \acc[L]^{K} \after \big(f^{L}\big)^{K} \after \delta[L]^{K} 
    \after \delta[K]
\\
& = &
\acc[K\cdot L] \after \concat_{K} \after \big(f^{L}\big)^{K} \after \delta[L]^{K} 
    \after \delta[K]
\\
& = &
\acc[K\cdot L] \after f^{K\cdot L} \after \concat_{K} \after \delta[L]^{K} 
    \after \delta[K]
\\
& = &
\acc[K\cdot L] \after f^{K\cdot L} \after \delta[K\cdot L] 
\\
& = &
\multinomial[K\cdot L](f).
\end{array} \]

\noindent For the convolution property in the first diagram in the
third row:
\[ \begin{array}{rcl}
+ \, \after \big(\multinomial[K](f) \otimes \multinomial[L](f)\big)
   \after \delta
& = &
\acc \after \concat \after 
   \big((\arr\after\multinomial[K](f)) \otimes 
   (\arr\after\multinomial[L](f))\big)
   \after \delta
\\
& = &
\acc \after \concat \after 
   \big((f^{K}\after\delta[K]) \otimes (f^{L}\after\delta[L])\big)
   \after \delta
\\
& = &
\acc \after f^{K+L} \after \concat \after 
   \big(\delta[K] \otimes \delta[L]\big) \after \delta
\\
& = &
\acc \after f^{K+L} \after \delta[K\!+\!L]
\\
& = &
\multinomial[K\!+\!L](f).
\end{array} \]

\noindent Finally, along the same lines:
\[ \begin{array}[b]{rcl}
\mzip \after \big(\multinomial[K](f) \otimes \multinomial[K](g)\big)
& = &
\acc \after \zip \after 
   \big((\arr\after\multinomial[K](f)) \otimes 
   (\arr\after\multinomial[K](g))\big)
\\
& = &
\acc \after \zip \after 
   \big((f^{K}\after\delta[K]) \otimes (g^{K}\after\delta[K])\big)
\\
& = &
\acc \after (f\otimes g)^{K} \after \zip \after 
   \big(\delta[K] \otimes \delta[K]\big)
\\
& = &
\acc \after (f\otimes g)^{K} \after \delta[K]
\\
& = &
\multinomial[K](f\otimes g).
\end{array} \eqno{\QEDbox} \]
\end{myproof}

We turn to the hypergeometric case. Proofs of the following results
are easy, since we have aready done the heavy-lifting earlier.

\begin{theorem}
\label{HypGeomThm}
The following diagrams about hypergeometric maps commute.
\[ \xymatrix@C-0.5pc@R-0.8pc{
\Mlt[L](Y)\ar[rr]^-{\hypergeometric[L,K]} & & \Mlt[K](Y)
& \qquad &
\Mlt[L](X)\ar[rr]^-{\hypergeometric[L,K]}\ar@/_2ex/[dr]_{\flrn} & & 
   \Mlt[K](X)\ar@/^2ex/[dl]^{\flrn}
\\
& X\ar@/^2ex/[ul]^{\multinomial[L](f)}\ar@/_2ex/[ur]_{\multinomial[K](f)} &
& &
& X &
} \]
\[ \xymatrix@C+1pc@R-0.8pc{
\Mlt[L](X)\otimes\Mlt[L](Y)\ar[d]_{\mzip}
   \ar[rr]^-{\hypergeometric[L,K]\otimes\hypergeometric[L,K]}
   & & \Mlt[K](X)\otimes\Mlt[K](Y)\ar[d]^{\mzip}
\\
\Mlt[L](X\otimes Y)\ar[rr]^-{\hypergeometric[L,K]} & & \Mlt[K](X\otimes Y)
} \]



\end{theorem}

\begin{myproof}
Commutation of the first triangle, on the left, follows directly from
the definition of $\hypergeometric[L,K]$, using the commutation of
multinomials with draw-and-delete in Theorem~\ref{MulnomThm}. Via
iterated application of the diagram on the left in
Proposition~\ref{DrawDeleteProp} one gets commutation of the second
triangle, on the right. For the rectangle we use
Proposition~\ref{MzipProp}~\eqref{MzipPropDD}. \QED
\end{myproof}

\section{Concluding remarks}

This paper contains some basic handwork in categorical probability,
introducing multisets as quotients, with associated multinomial and
hypergeometric distributions. It builds on and extends the development
of probability theory in Markov categories.

We have not included tensors of multisets, as operation
$\Mlt[K](X)\times \Mlt[L](Y) \rightarrow \Mlt[K\cdot L](X\otimes Y)$.
It is possible to define such an operation, via strength $\st
\coloneqq \zip \after (\delta[L]\otimes\idmap) \colon X\otimes Y^{L}
\rightarrow (X\times Y)^{L}$ for sequences. When one assumes that
coequalisers are preserved by tensors $\otimes$, one can define
strength for multisets $\mst \colon X \otimes \Mlt[L](Y) \rightarrow
\Mlt[L](X\otimes Y)$ such that $\mst \after (\idmap\otimes\acc) = \acc
\after \st$. Although strength for sequences is not commutative, this
strength for multisets does satisfy commutativity, in a suitably
graded sense. However, the problem is that these strengths, for
sequences and for multisets, are not natural, since they involve
copying. This generalises the findings in~\cite{Jacobs21b} that
tensors of multisets are not well-behaved in a probabilistic setting
and that the multizip operation should be used instead --- for
instance because it makes the (fixed-size) multiset functor monoidal
and commutes well with multinomial and hypergeometric distributions,
as shown here. However, not all is well with multizip, since it does
not make $\Mlt[K]$ into a \emph{monoidal} graded monad. Calculation of
a counterexample is quite intimidating.

It remains an interesting question, now with more urgency, what is
required to represent other discrete and also continuous distributions
in Markov categories.


\begin{thebibliography}{10}

\bibitem{AbramskyC09}
S.~Abramsky and B.~Coecke.
\newblock A categorical semantics of quantum protocols.
\newblock In K.~Engesser, {Dov}~M. Gabbay, and D.~Lehmann, editors, {\em
  Handbook of Quantum Logic and Quantum Structures: Quantum Logic}, pages
  261--323. North-Holland, Elsevier, Computer Science Press, 2009.
\newblock \href {https://doi.org/10.1016/b978-0-444-52869-8.50010-4}
  {\path{doi:10.1016/b978-0-444-52869-8.50010-4}}.

\bibitem{AdamekV08}
J.~Ad{\'a}mek and J.~Velebil.
\newblock Analytic functors and weak pullbacks.
\newblock {\em Theory and Appl. of Categories}, 21(11):191--209, 2008.

\bibitem{ChoJ19}
K.~Cho and B.~Jacobs.
\newblock Disintegration and {Bayesian} inversion via string diagrams.
\newblock {\em Math. Struct. in Comp. Sci.}, 29(7):938--971, 2019.
\newblock \href {https://doi.org/10.1017/s0960129518000488}
  {\path{doi:10.1017/s0960129518000488}}.

\bibitem{ChoJWW15b}
K.~Cho, B.~Jacobs, A.~Westerbaan, and B.~Westerbaan.
\newblock An introduction to effectus theory.
\newblock see \url{https://arxiv.org/abs/1512.05813}, 2015.

\bibitem{ClercDDG17}
F.~Clerc, F.~Dahlqvist, V.~Danos, and I.~Garnier.
\newblock Pointless learning.
\newblock In J.~Esparza and A.~Murawski, editors, {\em Foundations of Software
  Science and Computation Structures}, number 10203 in Lect. Notes Comp. Sci.,
  pages 355--369. Springer, Berlin, 2017.
\newblock \href {https://doi.org/10.1007/978-3-662-54458-7_21}
  {\path{doi:10.1007/978-3-662-54458-7_21}}.

\bibitem{CoeckeHK14}
B.~Coecke, C.~Heunen, and A.~Kissinger.
\newblock Categories of quantum and classical channels.
\newblock {\em Quantum Information Processing}, pages 1–--31, 2014.
\newblock \href {https://doi.org/10.1007/s11128-014-0837-4}
  {\path{doi:10.1007/s11128-014-0837-4}}.

\bibitem{CoeckeK16}
B.~Coecke and A.~Kissinger.
\newblock {\em Picturing Quantum Processes. A First Course in Quantum Theory
  and Diagrammatic Reasoning}.
\newblock Cambridge Univ. Press, 2016.
\newblock \href {https://doi.org/10.1017/9781316219317}
  {\path{doi:10.1017/9781316219317}}.

\bibitem{CoeckeS12}
B.~Coecke and R.~Spekkens.
\newblock Picturing classical and quantum {Bayesian} inference.
\newblock {\em Synthese}, 186(3):651--696, 2012.
\newblock \href {https://doi.org/10.1007/s11229-011-9917-5}
  {\path{doi:10.1007/s11229-011-9917-5}}.

\bibitem{CulbertsonS14}
J.~Culbertson and K.~Sturtz.
\newblock A categorical foundation for {Bayesian} probability.
\newblock {\em Appl. Categorical Struct.}, 22(4):647--662, 2014.
\newblock \href {https://doi.org/10.1007/s10485-013-9324-9}
  {\path{doi:10.1007/s10485-013-9324-9}}.

\bibitem{DahlqvistDG16}
F.~Dahlqvist, V.~Danos, and I.~Garnier.
\newblock Robustly parameterised higher-order probabilistic models.
\newblock In J.~Desharnais and R.~Jagadeesan, editors, {\em Int. Conf. on
  Concurrency Theory}, volume~59 of {\em LIPIcs}, pages 23:1--23:15. Schloss
  Dagstuhl, 2016.
\newblock \href {https://doi.org/10.4230/LIPIcs.CONCUR.2016.23}
  {\path{doi:10.4230/LIPIcs.CONCUR.2016.23}}.

\bibitem{DahlqvistK20}
F.~Dahlqvist and D.~Kozen.
\newblock Semantics of higher-order probabilistic programs with conditioning.
\newblock In {\em Princ. of Programming Languages}, pages 57:1--57:29. ACM
  Press, 2020.
\newblock \href {https://doi.org/10.1145/3371125} {\path{doi:10.1145/3371125}}.

\bibitem{DanosE11}
V.~Danos and T.~Ehrhard.
\newblock Probabilistic coherence spaces as a model of higher-order
  probabilistic computation.
\newblock {\em Information \& Computation}, 209(6):966--991, 2011.

\bibitem{DashS20}
S.~Dash and S.~Staton.
\newblock A monad for probabilistic point processes.
\newblock In D.~Spivak and J.~Vicary, editors, {\em Applied Category Theory
  Conference}, Elect. Proc. in Theor. Comp. Sci., 2020.
\newblock \href {https://doi.org/10.4204/EPTCS.333.2}
  {\path{doi:10.4204/EPTCS.333.2}}.

\bibitem{DashS21}
S.~Dash and S.~Staton.
\newblock Monads for measurable queries in probabilistic databases.
\newblock In A.~Sokolova, editor, {\em Math. Found. of Programming Semantics},
  2021.

\bibitem{Feller70I}
W.~Feller.
\newblock {\em An Introduction to Probability Theory and Its applications},
  volume~I.
\newblock Wiley, $3^{\textrm{rd}}$ rev. edition, 1970.
\newblock \href {https://doi.org/10.1063/1.3062516}
  {\path{doi:10.1063/1.3062516}}.

\bibitem{Fong12}
B.~Fong.
\newblock Causal theories: A categorical perspective on {Bayesian} networks.
\newblock Master's thesis, Univ.\ of Oxford, 2012.
\newblock see \url{https://arxiv.org/abs/1301.6201}.

\bibitem{Fritz20}
T.~Fritz.
\newblock A synthetic approach to {Markov} kernels, conditional independence,
  and theorems on sufficient statistics.
\newblock {\em Advances in Math.}, 370:107239, 2020.
\newblock \href {https://doi.org/10.1016/J.AIM.2020.107239}
  {\path{doi:10.1016/J.AIM.2020.107239}}.

\bibitem{FritzGPR20}
T.~Fritz, T.~Gonda, P.~Perrone, and E.~Rischel.
\newblock Representable {Markov} categories and comparison of statistical
  experiments in categorical probability.
\newblock See \url{https://arxiv.org/abs/2010.07416}, 2020.

\bibitem{FritzR20}
T.~Fritz and E.~Rischel.
\newblock Infinite products and zero-one laws in categorical probability.
\newblock {\em Compositionality}, 2(3), 2020.
\newblock \href {https://doi.org/10.32408/compositionality-2-3}
  {\path{doi:10.32408/compositionality-2-3}}.

\bibitem{FujiiKM16}
S.~Fujii, S.~Katsumata, and P.~Melli{\`{e}}s.
\newblock Towards a formal theory of graded monads.
\newblock In B.~Jacobs and C.~L{\"o}ding, editors, {\em Foundations of Software
  Science and Computation Structures}, number 9634 in Lect. Notes Comp. Sci.,
  pages 513--530. Springer, Berlin, 2016.
\newblock \href {https://doi.org/10.1007/978-3-662-49630-5_30}
  {\path{doi:10.1007/978-3-662-49630-5_30}}.

\bibitem{Jacobs15d}
B.~Jacobs.
\newblock New directions in categorical logic, for classical, probabilistic and
  quantum logic.
\newblock {\em Logical Methods in Comp. Sci.}, 11(3), 2015.
\newblock \href {https://doi.org/10.2168/lmcs-11(3:24)2015}
  {\path{doi:10.2168/lmcs-11(3:24)2015}}.

\bibitem{Jacobs16a}
B.~Jacobs.
\newblock Affine monads and side-effect-freeness.
\newblock In I.~Hasuo, editor, {\em Coalgebraic Methods in Computer Science
  (CMCS 2016)}, number 9608 in Lect. Notes Comp. Sci., pages 53--72. Springer,
  Berlin, 2016.
\newblock \href {https://doi.org/10.1007/978-3-319-40370-0_5}
  {\path{doi:10.1007/978-3-319-40370-0_5}}.

\bibitem{Jacobs18c}
B.~Jacobs.
\newblock From probability monads to commutative effectuses.
\newblock {\em Journ. of Logical and Algebraic Methods in Programming},
  94:200--237, 2018.
\newblock \href {https://doi.org/10.1016/j.jlamp.2016.11.006}
  {\path{doi:10.1016/j.jlamp.2016.11.006}}.

\bibitem{Jacobs21b}
B.~Jacobs.
\newblock From multisets over distributions to distributions over multisets.
\newblock In {\em Logic in Computer Science}. IEEE, Computer Science Press,
  2021.
\newblock See \url{https://arxiv.org/abs/2105.06908}.

\bibitem{Jacobs21c}
B.~Jacobs.
\newblock Learning from what's right and learning from what's wrong.
\newblock In A.~Sokolova, editor, {\em Math. Found. of Programming Semantics},
  2021.

\bibitem{Jacobs21a}
B.~Jacobs.
\newblock Multisets and distributions, in drawing and learning.
\newblock In A.~Palmigiano and M.~Sadrzadeh, editors, {\em Samson Abramsky on
  Logic and Structure in Computer Science and Beyond}. Springer, 2021, to
  appear.

\bibitem{JacobsZ21}
B.~Jacobs and F.~Zanasi.
\newblock The logical essentials of {Bayesian} reasoning.
\newblock In G.~Barthe, J.-P. Katoen, and A.~Silva, editors, {\em Foundations
  of Probabilistic Programming}, pages 295--331. Cambridge Univ. Press, 2021.
\newblock \href {https://doi.org/10.1017/9781108770750.010}
  {\path{doi:10.1017/9781108770750.010}}.

\bibitem{Joyal86}
A.~Joyal.
\newblock Foncteurs analytiques et esp\`eces de structures.
\newblock In G.~Labelle and P.~Leroux, editors, {\em Combinatoire Enumerative},
  number 1234 in Lect. Notes Math., pages 126--159. Springer, Berlin, 1986.
\newblock \href {https://doi.org/10.1007/bfb0072514}
  {\path{doi:10.1007/bfb0072514}}.

\bibitem{KellyL80}
M.~Kelly and M.~Laplaza.
\newblock Coherence for compact closed categories.
\newblock {\em Journ. of Pure \& Appl. Algebra}, 19:193--213, 1980.
\newblock \href {https://doi.org/10.1016/0022-4049(80)90101-2}
  {\path{doi:10.1016/0022-4049(80)90101-2}}.

\bibitem{Kock12}
A.~Kock.
\newblock Commutative monads as a theory of distributions.
\newblock {\em Theory and Appl. of Categories}, 26(4):97--131, 2012.

\bibitem{MiliusPS15}
S.~Milius, D.~Pattinson, and L.~Schr{\"{o}}der.
\newblock Generic trace semantics and graded monads.
\newblock In L.~Moss and P.~Sobocinski, editors, {\em Conference on Algebra and
  Coalgebra in Computer Science (CALCO 2015)}, volume~35 of {\em LIPIcs}, pages
  253--269. Schloss Dagstuhl, 2015.
\newblock \href {https://doi.org/10.4230/LIPIcs.CALCO.2015.253}
  {\path{doi:10.4230/LIPIcs.CALCO.2015.253}}.

\bibitem{OlmedoGKKM18}
F.~Olmedo, F.~Gretz, B.~Lucien Kaminski, J-P. Katoen, and A.~McIver.
\newblock Conditioning in probabilistic programming.
\newblock {\em ACM Trans. on Prog. Lang. \& Syst.}, 40(1):4:1--4:50, 2018.
\newblock \href {https://doi.org/doi.org/10.1145/3156018}
  {\path{doi:doi.org/10.1145/3156018}}.

\bibitem{Selinger07}
P.~Selinger.
\newblock Dagger compact closed categories and completely positive maps
  (extended abstract).
\newblock In P.~Selinger, editor, {\em Proceedings of the 3rd International
  Workshop on Quantum Programming Languages (QPL 2005)}, number 170 in Elect.
  Notes in Theor. Comp. Sci., pages 139--163. Elsevier, Amsterdam, 2007.
\newblock \href {https://doi.org/10.1016/j.entcs.2006.12.018}
  {\path{doi:10.1016/j.entcs.2006.12.018}}.

\bibitem{StatonYHKW16}
S.~Staton, H.~Yang, C.~Heunen, O.~Kammar, and F.~Wood.
\newblock Semantics for probabilistic programming: higher-order functions,
  continuous distributions, and soft constraints.
\newblock In {\em Logic in Computer Science}. IEEE, Computer Science Press,
  2016.
\newblock \href {https://doi.org/10.1145/2933575.2935313}
  {\path{doi:10.1145/2933575.2935313}}.

\end{thebibliography}

\appendix
\section{Appendix}\label{AppSec}

We sketch a proof of Proposition~\ref{MltCoprodProp}.  Using that
$\otimes$ distributes over $+$ we formulate the Binomial Theorem as a
`list-split' isomorphism $\lsplit$ in:
\begin{equation}
\label{ListSplitDiag}
\vcenter{\xymatrix{
(X+Y)^{K}\ar[rr]^-{\lsplit[K]}_-{\cong} & &
   \bigoplus\limits_{0\leq i\leq K} \binom{K}{i}\cdot 
   \Big(X^{i} \otimes Y^{K-i}\Big).
}}
\end{equation}

\noindent We use the dot $\cdot$ for copower, so that $n\cdot X = X +
\cdots + X$. The binomial coefficient $\binom{K}{i}$ occurs because
there are $\binom{K}{i}$ ways of turning a list of $X$'s of length $i$
and a list of $Y$'s of length $K-i$ into a list of $X+Y$'s of length
$K$, since the alternations of $X$ and $Y$ in $(X+Y)^{K}$ need to be
taken into account.

These $\lsplit$ isomorphisms in~\eqref{ListSplitDiag} are obtained by
induction on $K$. First, by definition,
\[ \begin{array}{rcccccccl}
(X+Y)^{0}
& \cong &
1
& \cong &
1\otimes 1
& \cong &
1\cdot \big(X^{0} \otimes Y^{0}\big)
& \cong &
\bigoplus\limits_{0\leq i\leq 0} \binom{0}{i}\cdot 
   \big(X^{i} \otimes Y^{0-i}\big).
\end{array} \]

\noindent Next, via the familiar argument, but now in categorical
form, using Pascal's identity:
\[ \begin{array}[b]{rcl}
(X+Y)^{K+1}
& \cong &
(X+Y)\otimes (X+Y)^{K}
\\[+0.2em]
& \cong &
X\otimes (X+Y)^{K} \,+\, Y \otimes (X+Y)^{K}
\\
& \cong &
X\otimes\left(\bigoplus\limits_{0\leq i\leq K} \binom{K}{i}\cdot 
   \big(X^{i} \otimes Y^{K-i}\big)\right) \,+\,
Y\otimes\left(\bigoplus\limits_{0\leq i\leq K} \binom{K}{i}\cdot 
   \big(X^{i} \otimes Y^{K-i}\big)\right)
\\
& \cong &
\left(\bigoplus\limits_{0\leq i\leq K} \binom{K}{i}\cdot 
   \big(X^{i+1} \otimes Y^{K-i}\big)\right) \,+\,
\left(\bigoplus\limits_{0\leq i\leq K} \binom{K}{i}\cdot 
   \big(X^{i} \otimes Y^{K+1-i}\big)\right)
\\[+1.2em]
& \cong &
\binom{K}{0}\cdot\big(X^{1}\otimes Y^{K}\big) + \cdots + 
   \binom{K}{K}\cdot\big(X^{K+1}\otimes Y^{0}\big) \,+\,
   \binom{K}{0}\cdot\big(X^{0}\otimes Y^{K+1}\big) + \cdots + 
   \binom{K}{K}\cdot\big(X^{K}\otimes Y^{1}\big)
\\[+0.2em]
& \cong &
\binom{K+1}{0}\cdot\big(X^{0}\otimes Y^{K+1}\big) \,+\,
\left(\bigoplus\limits_{1\leq i\leq K} 
   \left(\binom{K}{i-1}+\binom{K}{i}\right)\cdot 
   \big(X^{i} \otimes Y^{K+1-i}\big)\right) \,+\,
   \binom{K+1}{K+1}\cdot\big(X^{K+1}\otimes Y^{0}\big)
\\[+1.2em]
& \cong &
\bigoplus\limits_{0\leq i\leq K+1} \binom{K+1}{i}\cdot 
   \big(X^{i} \otimes Y^{K+1-i}\big).
\end{array} \]

A next step is to combine list-split with accumulation.

\begin{lemma}
\label{ListSplitAccLem}
For $K\geq 0$ we write $\accs[K]$ for the sum of cotuples of
accumulation maps in:
\[ \xymatrix@C+2pc{
\accs[K] \;\coloneqq\; 
   \Big(\bigoplus\limits_{0\leq i\leq K}\binom{K}{i}\cdot 
      \big(X^{i} \otimes Y^{K-i}\big)
    \ar[rr]^-{\bigoplus\limits_{0\leq i\leq K}\big[\acc[i]\otimes\acc[K-i]\big]} & &
      \bigoplus\limits_{0\leq i\leq K}\Mlt[i](X)\otimes\Mlt[K\!-\!i](Y)\Big)
} \]

\noindent Then:
\begin{enumerate}
\item $\accs[K] \after \lsplit = \bigoplus\limits_{0\leq i\leq K} 
   \Big((\acc[i]\otimes\acc[K\!-\!i]) \after \nabla\Big) \after \lsplit$;

\item $\accs[K] \after \lsplit \after \underline{\sigma} = \accs[K]
  \after \lsplit$ for each permutation $\sigma\in S_{K}$.
\end{enumerate}
\end{lemma}

\begin{myproof}
The first point says that that the the maps
$\acc[i]\otimes\acc[K\!-\!i]$ act the same on each of the
$\binom{K}{i}$-many alternations of $X$ and $Y$ in $(X+Y)^{K}$.  This
follows from an easy combinatorial argument. Similarly for the second
point. \QED
\end{myproof}

We are now in a position to define a multiset split map $\msplit$ in:
\begin{equation}
\label{MultSplitEqn}
\vcenter{\xymatrix@C-0.5pc@R-1.5pc{
(X+Y)^{K}\ar@/^2.8ex/[rr]\ar@/_2ex/[rr] & \vdots & 
    (X+Y)^{K}\ar@{->>}[rr]^-{\acc}\ar@/_1ex/[dr]_-{\lsplit} & & 
   \Mlt[K](X+Y)\ar@{-->}[dd]^-{\msplit}
\\
& & & \bigoplus\limits_{0\leq i\leq K}\binom{K}{i}\cdot 
      \big(X^{i} \otimes Y^{K-i}\big)\ar@/_1ex/[dr]_-{\accs[K]\quad} &
\\
& & & & \bigoplus\limits_{0\leq i\leq K}\Mlt[i](X)\otimes\Mlt[K\!-\!i](Y)
}}
\end{equation}

Our aim is to show that $\msplit$ is an isomorphism. There is an obvious
map in the reverse direction, which we already write as $\msplit^{-1}$
in anticipation of the proof. It's define via the sum of multisets
from Definition~\ref{SumDef}.
\begin{equation}
\label{MultSplitInvEqn}
\vcenter{\xymatrix@C+2.5pc@R-0.0pc{
\bigoplus\limits_{0\leq i\leq K}\Mlt[i](X)\otimes\Mlt[K\!-\!i](Y)
   \ar@/_2ex/[drr]_-{\msplit^{-1}\quad}
   \ar[rr]^-{\bigoplus\limits_{0\leq i\leq K}
   \Mlt[i](\kappa_{1})\otimes\Mlt[K\!-\!i](\kappa_{2})} & &
   \bigoplus\limits_{0\leq i\leq K}\Mlt[i](X+Y)\otimes\Mlt[K\!-\!i](X+Y)
   \ar[d]^-{[\,+\,]_{0\leq i\leq K}}
\\
& & \Mlt[K](X+Y)
}}
\end{equation}

\noindent It is now ``obvious'' that $\msplit$ and $\msplit^{-1}$ are
each other's inverses, proving
Proposition~\ref{MltCoprodProp}~\eqref{MltCoprodPropCoprod}.

We add a proof of
Proposition~\ref{MltCoprodProp}~\eqref{MltCoprodPropNum}, stating that
$\Mlt[K](\underline{n}) \cong \left(\binom{n}{K}\right)\cdot 1$, where
the multichoose coefficient is defined as $\left(\binom{n}{K}\right) =
\binom{n+K-1}{K}$. This result is obtained by induction on $n\geq 1$.
For $n = 1$ we get, by
Lemma~\ref{MultSpecialLem}~\eqref{MultSpecialLemFinal}:
\[ \begin{array}{rcccccccl}
\Mlt[K](\underline{1})
& = &
\Mlt[K](1)
& \cong &
1
& = &
\binom{1+K-1}{K}\cdot 1
& = &
\left(\binom{1}{K}\right)\cdot 1.
\end{array} \]

\noindent Next,
\[ \begin{array}{rcll}
\Mlt[K](\underline{n+1})
\hspace*{\arraycolsep}\cong\hspace*{\arraycolsep}
\Mlt[K](\underline{n}+1)
& \cong &
\displaystyle\bigoplus\limits_{0\leq i\leq K} 
   \Mlt[i](\underline{n})\otimes\Mlt[K\!-\!i](1)
   & \mbox{by Proposition~\ref{MltCoprodProp}~\eqref{MltCoprodPropCoprod}}
\\
& \cong &
\displaystyle\bigoplus\limits_{0\leq i\leq K} 
   \left[\left(\!\binom{n}{i}\!\right)\cdot 1\right] \otimes 1
   & \mbox{by induction hypothesis, and 
     Lemma~\ref{MultSpecialLem}~\eqref{MultSpecialLemFinal}}
\\[+1.5em]
& \cong &
\displaystyle\left[\sum_{0\leq i\leq K} \left(\!\binom{n}{i}\!\right)\right]
   \cdot 1
\\[+1.5em]
& = &
\displaystyle\left(\!\binom{n+1}{K}\!\right) \cdot 1.
\end{array} \]

\noindent The latter equation is a basic property of multichoose.

\end{document}